\documentclass[10pt, onecolumn, a4paper, fleqn]{article}

\usepackage{STY/PACKAGES}
\usepackage{STY/DEFINITIONS}
\usepackage{STY/SETTINGS}
\usepackage{STY/TIKZ}
\usepackage{STY/TITLEPAGE}

\usepackage{mdframed} 

\begin{document}
\edef\myindent{\the\parindent}
\author{Kronberg, Vì \and Anthonissen, Martijn \and ten Thije Boonkkamp, Jan \and IJzerman, Wilbert}
\title{Two-Dimensional Freeform Reflector Design with a Scattering Surface}

\thispagestyle{empty} 
\pagenumbering{gobble} 

\newgeometry{margin=19mm,top=30mm}

\begin{figure}[t]
	\centering
	\includegraphics[width=0.4\textwidth]{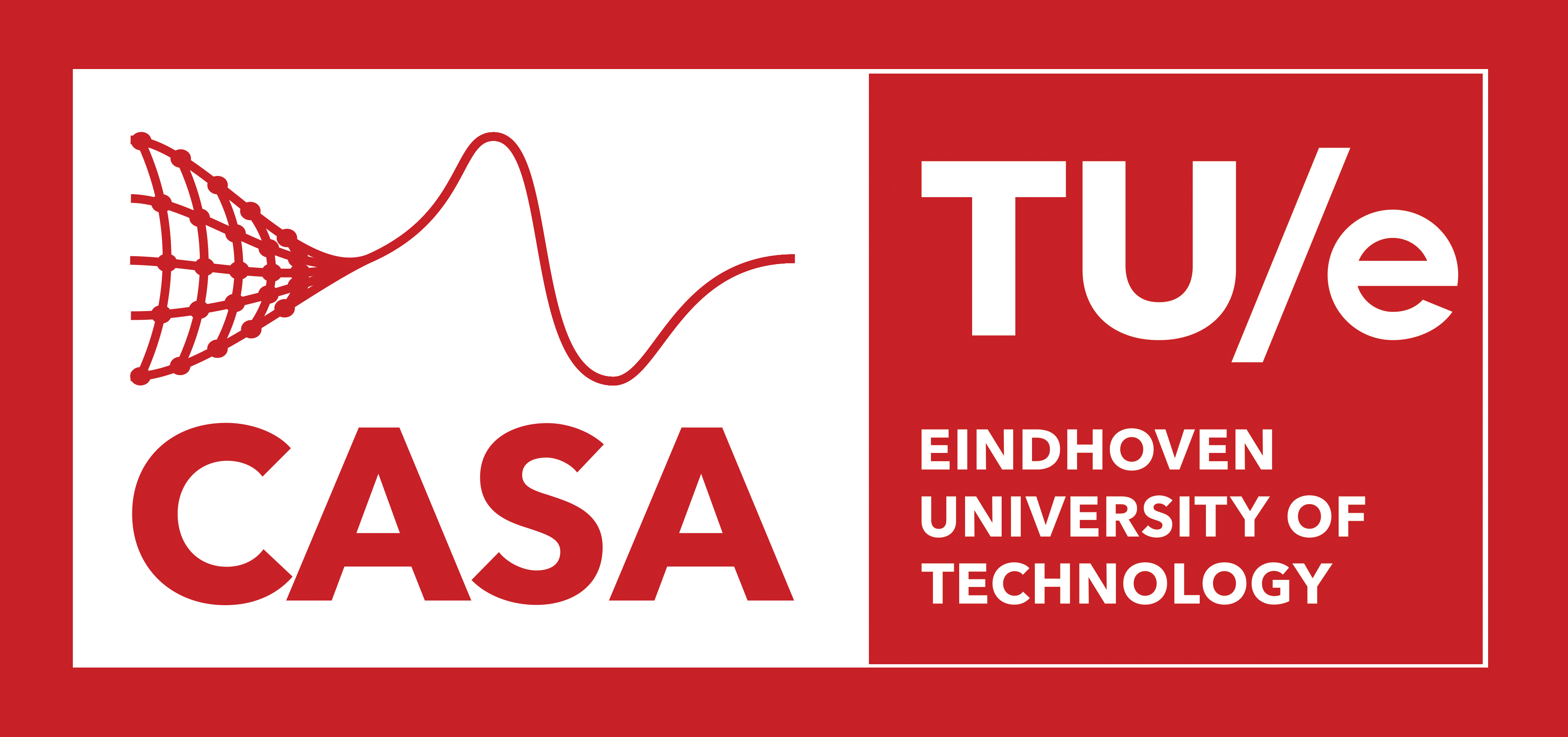}
\end{figure}

\vspace*{-0.7cm}
\begin{center}
	\LARGE \textbf{Two-Dimensional Freeform Reflector Design\\with a Scattering Surface}\\[0.5cm]
\end{center}

\begin{center}
	\large
	{\itshape V\`{i}}~\textsc{Kronberg},\textsuperscript{1,*} {\itshape Martijn}~\textsc{Anthonissen},\textsuperscript{1}\\
	{\itshape Jan}~\textsc{ten Thije Boonkkamp},\textsuperscript{1} and {\itshape Wilbert}~\textsc{IJzerman}\textsuperscript{1,2}
\end{center}

\begin{changemargin}{1.5cm}{1.5cm}
\noindent
\textsuperscript{1}{\itshape Department of Mathematics and Computer Science, Eindhoven University of Technology},\\PO Box 513, 5600 MB Eindhoven, The Netherlands\\
\textsuperscript{2}{\itshape Signify Research}, High Tech Campus 7, 5656 AE Eindhoven, The Netherlands\\
\textsuperscript{*}{\color{blue}\href{mailto:v.c.e.kronberg@tue.nl}{\texttt{v.c.e.kronberg@tue.nl}}}\\
\url{https://www.win.tue.nl/~martijna/Optics/}\\
\end{changemargin}

\hrule
\begin{changemargin}{1.5cm}{1.5cm}
{\textbf{\textsc{PACS:}} 02.30.Z $\cdot$ 42.15.−i $\cdot$ z42.25.Fx $\cdot$ 42.79.Fm\\[1mm]
\textbf{\textsc{AMS:}} 78A05 $\cdot$ 78A45 $\cdot$ 78A46}
\end{changemargin}
\hrule
\begin{changemargin}{1.5cm}{1.5cm}
	\textsc{\textbf{Abstract:}}
	We combine two-dimensional freeform reflector design with a scattering surface modelled using microfacets, i.e., small specular surfaces representing surface roughness.
	The model results in a convolution integral for the scattered light intensity distribution, which yields an inverse specular problem after deconvolution.
	Thus, the shape of a reflector with a scattering surface may be computed using deconvolution, followed by solving the typical inverse problem of specular reflector design.
\end{changemargin}
\hrule

\clearpage

\pagestyle{main}
\pagenumbering{arabic}
\setcounter{page}{1}
\restoregeometry

\section{Introduction}
The inverse problem of light transport within the field of computational illumination optics is concerned with computing an optical system --- typically lenses or reflectors --- such that a given source light distribution is transformed into a desired target distribution \cite{elmerOpticalDesignReflectors1980, olikerReconstructingReflectingSurface1989, prinsInverseMethodsIllumination2014, romijnGeneratedJacobianEquations2021, vanroosmalenDesignFreeformTworeflector2021}.
A common situation involves shaping light from a point source since this is the approximate distribution of a light-emitting diode (LED).
Romijn extensively studied shaping point source illumination using reflectors and lenses by constructing numerical solutions to the so-called generalised Monge-Amp{\`e}re equations \cite{romijnInverseReflectorDesign2020, romijnGeneratedJacobianEquations2021}.
For cylindrically and rotationally symmetric reflector problems, we can restrict our attention to two dimensions and solve two coupled ordinary differential equations (ODEs) to compute the specular surface \cite{kellerInverseScatteringProblem1959, maesMathematicalMethods2D1994}.

Since the introduction of LED light sources, there is now a strong demand for aesthetic lighting in terms of colour and light shaping.
Diffusive media or scattering surfaces can be utilised to homogenise light from an LED source or hide the LED itself  \cite[p.~24]{chavesIntroductionNonimagingOptics2016}.
Additionally, no actual reflector surface is a perfect mirror --- they all exhibit some form of light scattering due to surface defects introduced in the manufacturing process \cite[p.~24]{chavesIntroductionNonimagingOptics2016}.
Since the aforementioned methods of computing reflectors --- i.e., Monge-Amp{\`e}re equations and coupled ODEs --- only consider specular reflection from a perfect mirror, the effect of light scattering due to surface roughness is absent.
As such, real reflectors manufactured according to these designs may result in a broadened outgoing light distribution and blurring of finer features due to surface defects.
The present work aims to integrate surface light scattering into the design process of two-dimensional freeform reflectors in a mathematical way, thus allowing scattering surfaces to be utilised or detrimental effects due to surface defects to be pre-emptively mitigated.
Literature concerning the unification of scattering and freeform optics is minimal.
The best reference we have found is Lin \textit{et al.}, who designed a lens with a freeform scattering inner surface and a spherical outer one \cite{linNovelOpticalLens2015}.
The freeform surface was designed by an iterative optimisation procedure whereby a freeform surface represented by B{\'e}zier curves was first computed and then modified iteratively to take into account the difference between the prescribed target distribution and a raytraced one.

In contrast, we shall focus on computing two-dimensional freeform reflectors with a scattering surface using a direct inverse method.
Specifically, we shall derive a surface light scattering model by endowing a smooth curve (representing the surface) with microscopic roughness using so-called \textit{microfacets} and consider how the outgoing light distribution changes by their orientations.
The resulting expression for the scattered light is a convolution integral between a probability density function dictating the microfacets orientations (and is hence related to the surface roughness) and a so-called \textit{virtual} specular light distribution representing the reflected light from a perfectly smooth reflector.
Thus, assuming the roughness of the surface can be estimated \textit{a priori}, deconvolution can be utilised to compute a virtual specular light distribution, which can then be inserted as the target distribution in the inverse two-dimensional specular reflector problem solved by Maes \cite{maesMathematicalMethods2D1994}.

The structure of the manuscript is as follows.
First, the theory needed to develop the surface scattering model is introduced (zero-{\`e}tendue sources --- Sec.~\ref{sec:etendue}, the bidirectional reflectance distribution function --- Sec.~\ref{sec:BRDF}, and the microfacets --- Sec.~\ref{sec:microfacets}).
Then, the surface scattering model is derived for parallel-ray and point sources in two dimensions (Sec.~\ref{sec:convolution}), and the necessary two-dimensional specular design ODEs are derived (Secs.~\ref{sec:specParallel} and \ref{sec:specPoint}).
Next, several aspects of the model are verified using raytracing by considering two numerical examples (Sec.~\ref{sec:numerical}), and finally, some conclusions and proposed extensions are presented in Sec.~\ref{sec:conclusion}.

\section{Scattering Model and Freeform Reflector Design}
\noindent This section treats the theoretical aspects needed to develop the microfacet scattering model and to apply it in the context of two-dimensional freeform reflector design.

\subsection{Key Assumptions}
We have made several assumptions throughout our derivation of the convolution integral governing scattering in our model.
The key assumptions are discussed here, and any additional assumptions will be introduced when they become relevant.
The first assumption is that light scattering can be described using geometric optics by considering incoming, specularly reflected and scattered rays. Statistically, this is equivalent to the more physical notion of one incident direction yielding a cone of outgoing directions.
Additionally, light scattering is assumed to be fully elastic, i.e., the incident energy is scattered without absorption or other losses.
The medium surrounding our reflector is also assumed to be lossless, and light is assumed to be scattered exclusively due to surface roughness.
Finally, the addition of scattering in the system is assumed not to break the two-dimensional nature of the problem, i.e., we consider in-plane scattering.

\subsection{Geometry}
\begin{figure}[H]
	\centering
	\includegraphics[width=0.4\linewidth]{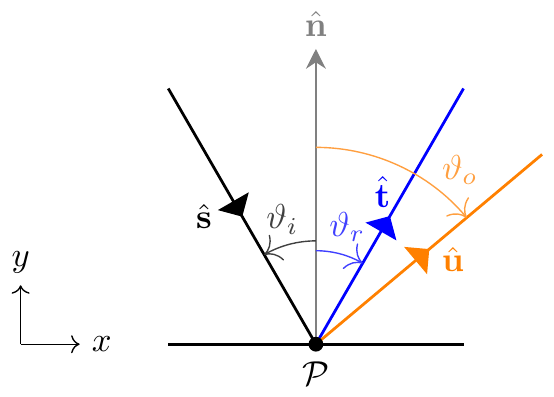}
	\captionsetup{width=0.8\linewidth}
	\caption{Geometry of scattering; $\th_r = -\th_i, \th_o < 0$ and $\th_i > 0$.}
	\label{fig:geometry}
\end{figure}

\noindent The two-dimensional geometry of scattering used throughout this manuscript is outlined in Fig.~\ref{fig:geometry}.

\noindent Here, we see an incident ray along direction $\us$ striking the surface at a point $\mathcal{P}$, resulting in a specular ray along $\ut$ for perfect specular reflection or a scattered ray along $\uu$.
The directions $\us$, $\ut$ and $\uu$ are given by angles $\th_i \in (-\pi/2,\pi/2)$, $\th_r = -\th_i$ and $\th_o \in (-\pi/2,\pi/2)$, respectively, i.e.,
\begin{equation}
	\us = \big(\!\sin(\th_i),-\cos(\th_i)\big)^\intercal, \quad \ut = \big(\!-\sin(\th_r),\cos(\th_r)\big)^\intercal, \quad \uu = \big(\!-\sin(\th_o),\cos(\th_o)\big)^\intercal.
\end{equation}

\subsection{Scattering Model}
\noindent Having introduced the geometry, let us now consider how the model is constructed.
We shall first introduce the concept of zero-{\'e}tendue sources, followed by the bidirectional reflectance distribution function (BRDF) and, finally, the microfacets used to model surface roughness.

\subsubsection{Zero-{\'e}tendue Sources}\label{sec:etendue}
The concept of {\'e}tendue is easiest to conceptualise in phase space --- the space in which all possible states of the system are represented.
The phase space of our two-dimensional geometry consists of two dimensions --- one spatial coordinate $x$ representing the position and one coordinate representing the momentum $p := n \sin(\th)$ of the ray at $x$ in direction $\th$, where $n$ is the refractive index.
We shall always take $n$ as unity, so that $p \equiv \sin(\th)$.

\begin{figure}[H]
	\centering
	\includegraphics[width=0.45\linewidth,page=2]{TikZ/sources/sources}\hfill
	\includegraphics[width=0.45\linewidth,page=1]{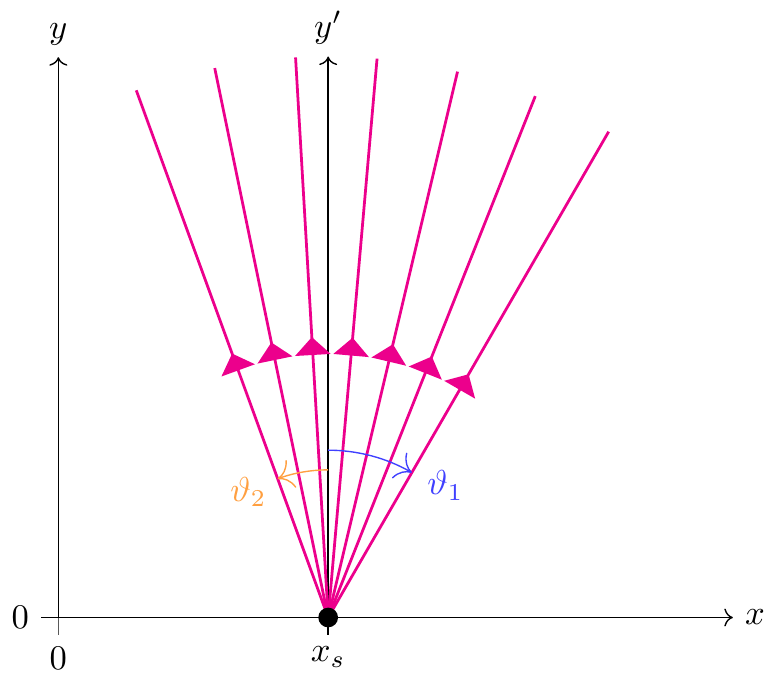}
	\captionsetup{width=\linewidth}
	\caption{Illustration of a parallel source of length $\ell_s$ emitting rays in direction $\th = \th_s$ (left) and a point source at $x = x_s$ emitting rays in directions $\th \in [\th_1,\th_2]$ (right); $\th_1$, $\th_s < 0$ and $\th_2 > 0$.}
	\label{fig:sources}
\end{figure}

The {\'e}tendue in our two-dimensional phase space is given by, cf.~\cite[Eq.~(5.50), p.~89]{boydRadiometryDetectionOptical1983},
\begin{equation}\label{eq:etendue_ray}
	\mathcal{U} = \int_{\mathcal{X}} \int_{\Theta} \cos(\th) \, \dd \th \dd x,
\end{equation}
where $\mathcal{X}$ and $\Theta$ are the spatial and angular domains of integration, respectively.
We shall consider two-dimensional parallel-ray sources (referred to as \textit{parallel sources} henceforth) and point sources in this manuscript.
In the case of parallel sources, the rays are emitted in a fixed direction $\th_s$, modulated along the spatial extent $\ell_s$ of the source, whilst in the case of point sources, the source is located at a fixed position $x_s$, where it emits modulated light in all directions $\th$.
The sources are illustrated in Fig.~\ref{fig:sources}, where the point source has been restricted to emitting rays in the interval $\th \in [\th_1,\th_2] \subseteq (-\pi/2,\pi/2)$.
The \textit{radiances} $L$ [W $\cdot$ rad$^{-1}$ $\cdot$ m$^{-1}$] of these sources are summarised in the definitions below.

\begin{definition}\label{def:parallelSource}
	The \textbf{radiance} of a two-dimensional parallel source emitting rays in direction $\th_s$ is given by
	\begin{equation}
		L(x,\th) = M(x) \delta(\th - \th_s),
	\end{equation}
	where $M(x)$ \textnormal{[W} $\cdot$ \textnormal{m}$^{-1}$\textnormal{]} is the \textbf{exitance} of the source at position $x$ and $\delta(\th - \th_s)$ \textnormal{[rad}$^{-1}$\textnormal{]} is the Dirac delta function.
\end{definition}
\begin{definition}\label{def:pointSource}
	The \textbf{radiance} of a two-dimensional point source at position $x_s$ is given by
	\begin{equation}
		L(x,\th) = I(\th) \delta(x - x_s),
	\end{equation}
	where $I(\th)$ \textnormal{[W} $\cdot$ \textnormal{rad}$^{-1}$\textnormal{]} is the \textbf{intensity} of the source in direction $\th$ and $\delta(x - x_s)$ \textnormal{[m}$^{-1}$\textnormal{]} is the Dirac delta function.
\end{definition}

\begin{figure}[H]
	\centering
	\includegraphics[width=0.48\linewidth,page=4]{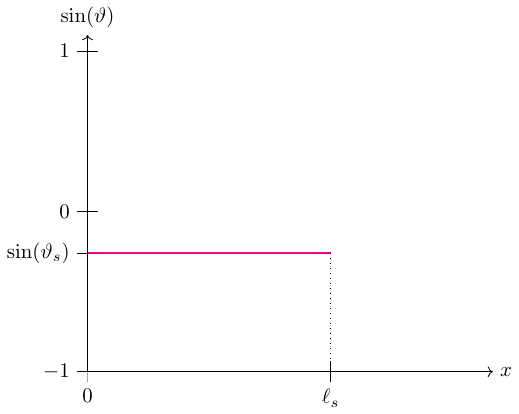}\hfill
	\includegraphics[width=0.48\linewidth,page=3]{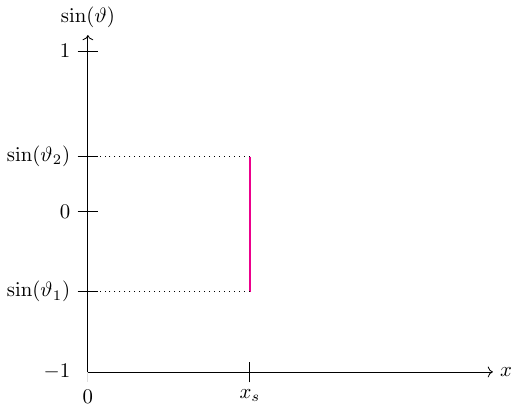}
	\captionsetup{width=\linewidth}
	\caption{The nonvanishing regions of phase space for the parallel (left) and point (right) sources in Fig.~\ref{fig:sources}.}
	\label{fig:sourcesPS}
\end{figure}

\noindent The nonvanishing regions of phase space for the parallel and point sources in Fig.~\ref{fig:sources} are shown in Fig.~\ref{fig:sourcesPS}, i.e., a horizontal and vertical line segment, respectively.
Returning to the definition of {\'e}tendue, Eq.~\eqref{eq:etendue_ray}, it is clear that it vanishes for both these sources since they form line segments of zero width in phase space ---  the parallel source has vanishing angular extent, and the point source has vanishing spatial extent.
Hence, they are so-called \textit{zero-{\'e}tendue sources} --- the only type of light sources we shall consider in this manuscript.

\subsubsection{Bidirectional Reflectance Distribution Function}\label{sec:BRDF}
\noindent The \textit{bidirectional reflectance distribution function} (BRDF) $B(\th_i,\th_o)$ [rad$^{-1}$] was first formulated by Nicodemus in 1965 \cite{nicodemusDirectionalReflectanceEmissivity1965}, and it describes the way light from direction $\th_i$ is scattered from the surface into direction $\th_o$ \cite{harveyUnderstandingSurfaceScatter2019} (recall Fig.~\ref{fig:geometry}).
To see how, let $L_i(\th_i)$ [W $\cdot$ rad$^{-1}$ $\cdot$ m$^{-1}$] be the incoming radiance from direction $\th_i$, and let $L_o(\th_o)$ [W $\cdot$ rad$^{-1}$ $\cdot$ m$^{-1}$] be the outgoing (i.e., scattered) radiance in direction $\th_o$.
Then, the so-called \textit{scattering equation}, yielding the outgoing radiance in direction $\th_o$, may be formulated as, cf.~\cite[Eq.~(7)]{nicodemusDirectionalReflectanceEmissivity1965},
\begin{equation}\label{eq:scatteringEq}
	L_o(\th_o) = \int_{-\pi/2}^{\pi/2} B(\th_i,\th_o) L_i(\th_i) \cos(\th_i) \, \dd\th_i.
\end{equation}

It can be shown that any physical BRDF must satisfy the following properties \cite{duvenhageNumericalVerificationBidirectional2013}:
\begin{itemize}
	\item Positivity:
	\begin{equation}\label{eq:positivity}
		\forall \th_i, \th_o \in (-\pi/2,\pi/2)\!: B(\th_i,\th_o) \geq 0,
	\end{equation}
	which must hold for obvious reasons --- both $L_i$ and $L_o$ are nonnegative in Eq.~\eqref{eq:scatteringEq}, so naturally, this applies to the BRDF as well.
	\item Helmholtz reciprocity:
	\begin{equation}\label{eq:HelmholtzReciprocity}
		B(\th_i,\th_o) = B(\th_o,\th_i),
	\end{equation}
	which encapsulates the intuitive fact that the flux in direction $\th_o$ from a source at $\th_i$ is the same as the flux in direction $\th_i$ from \textit{the same source} at $\th_o$, assuming the rest of the system remains the same.
	\item Energy conservation:
	\begin{equation}\label{eq:energyConservation}
		\forall \th_i \in (-\pi/2,\pi/2)\!: \int_{-\pi/2}^{\pi/2} B(\th_i,\th_o) \cos(\th_o) \, \dd\th_o = 1,
	\end{equation}
	which represents conservation of flux, implying that all the incoming light must be scattered in some direction.
	Note that this could be less than unity if absorbance or other losses were considered, but these effects are outside this work's scope.
\end{itemize}

\paragraph{Specular reflection as a BRDF}

\begin{figure}[H]
	\centering
	\includegraphics[width=0.35\linewidth]{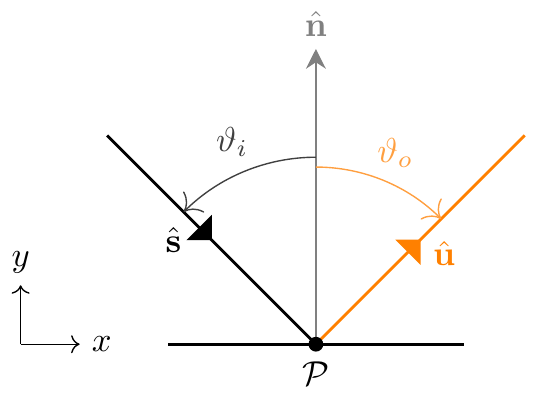}
	\captionsetup{width=0.8\linewidth}
	\caption{Perfect specular reflection from a flat surface; $\th_o = -\th_i < 0$ and $\th_i > 0$.}
	\label{fig:LoR}
\end{figure}

For example, let us consider the BRDF of specular reflection from a (locally) flat surface --- see Fig.~\ref{fig:LoR}.
The familiar law of reflection yields $\th_o \equiv \th_r = - \th_i$ since the angles originate from the unit surface normal $\un$ at the point of intersection.
This fact leads to the following definition for the BRDF of specular reflection.

\begin{definition}\label{def:specBRDF}
	The bidirectional reflectance distribution function (BRDF) of specular reflection is given by
	\begin{equation}\label{eq:specBRDF}
		B(\th_i,\th_o) = \frac{\delta(\th_i + \th_o)}{\cos(\th_o)},
	\end{equation}
	where $\delta$ is the Dirac delta function.
\end{definition}

\noindent This BRDF fulfils:
\begin{itemize}
	\item Positivity by the properties of the Dirac delta function.
	\item Helmholtz reciprocity, Eq.~\eqref{eq:HelmholtzReciprocity}:
	\begin{equation}
		B(\th_i,\th_o) = \frac{\delta(\th_i + \th_o)}{\cos(\th_o)} = \frac{\delta(\th_o + \th_i)}{\cos(\th_i)} = B(\th_o,\th_i),
	\end{equation}
	since the Dirac delta function is nonzero only when $\th_o = - \th_i$, and cosine is even.
	\item Energy conservation, Eq.~\eqref{eq:energyConservation}. Let $\th_i \in (-\pi/2,\pi/2)$, then
	\begin{equation}
		\int_{-\pi/2}^{\pi/2} \delta(\th_i + \th_o) \, \dd\th_o = 1,
	\end{equation}
	by the definition of the Dirac delta function.
\end{itemize}
Additionally, the scattering equation, Eq.~\eqref{eq:scatteringEq}, yields
\begin{equation}\label{eq:LoSpec}
	\begin{split}
		L_o(\th_o) \cos(\th_o) = \int_{-\pi/2}^{\pi/2} \delta(\th_i + \th_o) L_i(\th_i) \cos(\th_i) \, \dd\th_i &= L_i(-\th_o) \cos(-\th_o)\\
		&= L_i(\th_i) \cos(\th_i).
	\end{split}
\end{equation}
Note that we are keeping the cosine terms in this relation for later, but note that we could write $L_o(\th_o) = L_i(\th_i)$.
That is, the radiance in the outgoing direction $\th_o$ is exactly the radiance from the incoming direction $\th_i$, which is what we expect for specular reflection.

\subsubsection{Microfacets}\label{sec:microfacets}
\noindent Having treated specular reflection, we are now ready to develop a simple surface roughness model based on the concept of microfacets, i.e., small, tilted, specular sections superimposed on the macroscopic reflector.
By describing the orientations of the microfacets along the reflector surface using some suitable probability density function, a realistic model for surface roughness may be constructed.
This concept has been used extensively in computer-generated imagery (CGI) \cite[Ch.~8.4]{pharrPhysicallyBasedRendering2017}.
Of particular interest is the microfacet model developed by Torrance and Sparrow in 1967 \cite{torranceTheoryOffSpecularReflection1967}, and later applied to CGI by Cook and Torrance in 1982 \cite{cookReflectanceModelComputer1982}.
We have opted to develop a simpler model inspired by the one introduced by Torrance and Sparrow (in particular, we will not consider effects such as shadowing and masking) since our primary goal is to unify surface scattering with inverse design, starting with as few complexities as possible whilst remaining realistic.

\begin{figure}[ht!]
	\centering
	\includegraphics[width=0.7\linewidth]{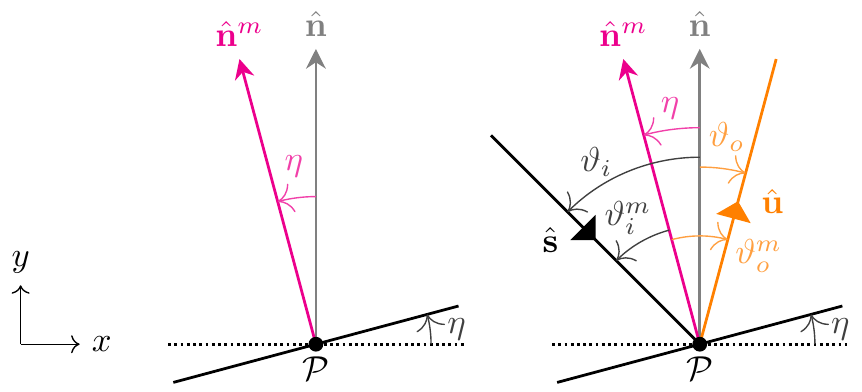}
	\captionsetup{width=0.8\linewidth}
	\caption{The geometry of a microfacet (solid black line) versus the macroscopic surface (dotted black line); $\th_o, \th_o^m < 0$ and $\eta, \th_i, \th_i^m > 0$.}
	\label{fig:microfacets}
\end{figure}

To describe the microfacets, consider zooming in sufficiently close so we can observe a locally flat section of the rough reflector.
Next, rotate the view such that the unit normal of the reflector, $\un$, aligns with the $y$-axis; this yields the view in Fig.~\ref{fig:LoR} and defines our coordinate system with its origin at $\mathcal{P}$.
Finally, consider tilting this locally flat reflector piece by some $\eta \in (-\pi/2,\pi/2)$, whilst keeping the angle of the incident light constant with respect to the original unit normal $\un$ --- see Fig.~\ref{fig:microfacets}, where $\th_i^m$ and $\th_o^m$ are the incident and outgoing directions with respect to the microfacet normal $\un^m$, respectively.
Mathematically, we define a rotation matrix
\begin{equation}\label{eq:rotMatrix}
	\mathbf{R}(\theta) :=
	\begin{pmatrix}
		\cos(\theta) & -\sin(\theta)\\
		\sin(\theta) & \cos(\theta)
	\end{pmatrix},
\end{equation}
to construct the microfacet normal $\un^m := \mathbf{R}(\eta)\un$.

Since the microfacet is specular, the law of reflection yields $\th_o^m \equiv \th_r^m = -\th_i^m$.
Using this, together with the geometry in Fig.~\ref{fig:microfacets}, we get
\begin{equation}\label{eq:microfacetRelations}
	\th_o^m = -\th_i^m, \quad \th_i = \th_i^m + \eta, \quad \th_o = \th_o^m + \eta \quad \Longrightarrow \quad \th_i = -\th_o + 2\eta.
\end{equation}
That is, the outgoing direction with respect to the original macroscopic normal is the negative incident direction, plus two times the perturbation of the normal.
Inserting this expression for $\th_i$ in Eq.~\eqref{eq:LoSpec} yields
\begin{equation}\label{eq:L_o-microfacets1}
	L_o(\th_o) \cos(\th_o) = L_i(-\th_o + 2\eta) \cos(-\th_o + 2\eta).
\end{equation}
This relation gives the outgoing radiance $L_o$ in terms of the incoming radiance $L_i$ for a microfacet with orientation $\eta$ with respect to the macroscopic surface normal.

\paragraph{Distributed orientations}
To use microfacets to describe scattering from a surface, rather than fixing $\eta$, we shall consider that it is sampled from some distribution related to the roughness properties of the surface.
That is, we do not know \textit{a priori} which microfacet orientation we strike, but rather we wish to describe the distribution of possible orientations using some probability density function $\tilde{p}(\eta)$ with support $\supp(\tilde{p}) = [\eta_1, \eta_2] \subseteq (-\pi/2,\pi/2)$, i.e., $\int_{\eta_1}^{\eta_2} \tilde{p}(\eta) \, \dd\eta = 1$.
Note that whilst the support of $\tilde{p}$ can formally extend to $(-\pi/2,\pi/2)$, we will typically consider situations where the significant contributions are concentrated in a much more narrow region centred around $\eta = 0$.
This shape is consistent with measured BRDFs from machined surfaces \cite[Ch.~4]{stoverOpticalScatteringMeasurement2012}.

To get the outgoing radiance, we now multiply the right-hand side of Eq.~\eqref{eq:L_o-microfacets1} by $\tilde{p}(\eta)$ and integrate over all valid microfacet orientations, leading to the below definition.

\begin{definition}\label{def:microfacetOutgoing}
	Let $\eta \in [\eta_1, \eta_2] \subseteq (-\pi/2,\pi/2)$ be the microfacet orientation with respect to the macroscopic normal $\un$, sampled from the probability density function $\tilde{p}(\eta)$.
	Let $L_o$ be the outgoing (i.e., scattered) radiance in direction $\th_o$ due to an incoming radiance $L_i$ from direction $\th_i$.
	Then:
	\begin{equation}
		L_o(\th_o) \cos(\th_o) = \int_{\eta_1}^{\eta_2} \tilde{p}(\eta) L_i(-\th_o + 2\eta) \cos(-\th_o + 2 \eta) \, \dd\eta.
	\end{equation}
\end{definition}

\noindent By using $\eta \equiv (\th_i+\th_o)/2$ --- recall Eq.~\eqref{eq:microfacetRelations} --- and changing the integration variable to $\th_i$, we get
\begin{equation}
	L_o(\th_o) \cos(\th_o) = \frac{1}{2} \int_{2\eta_1 - \th_o}^{2\eta_2 - \th_o} \tilde{p}\left(\frac{\th_i+\th_o}{2}\right) L_i(\th_i) \cos(\th_i) \, \dd\th_i.
\end{equation}
Since $\th_o \in (-\pi/2,\pi/2)$, and $\supp(\tilde{p}) = [\eta_1,\eta_2] \subseteq (-\pi/2, \pi/2)$, we can rewrite this integral as
\begin{equation}\label{eq:refPoint}
	L_o(\th_o) \cos(\th_o) = \frac{1}{2} \int_{-\pi/2}^{\pi/2} \tilde{p}\left(\frac{\th_i+\th_o}{2}\right) L_i(\th_i) \cos(\th_i) \, \dd\th_i.
\end{equation}
Note that the above expression does not consider whether a ray can physically reach $\th_o$ from $\th_i$.
As an extreme example, consider near-grazing incident light, where any nonzero perturbation $\eta$ of the normal could mean the incident ray strikes the back of the microfacet.
Implicit in the derivation of this model is that such unphysical situations do not occur.

Next, let
\begin{equation}
	p(\th_i+\th_o) := \tilde{p}\left(\frac{\th_i+\th_o}{2}\right)\!/2.
\end{equation}
Before proceeding, let us briefly verify that $p$ is also a PDF.
Let $\alpha := \th_i + \th_o = 2\eta$, then
\begin{equation}
	1 = \int_{\eta_1}^{\eta_2} \tilde{p}(\eta) \, \dd\eta
	= \frac{1}{2} \int_{\alpha_1}^{\alpha_2} \tilde{p}\left(\frac{\alpha}{2}\right) \,\dd\alpha
	= \int_{\alpha_1}^{\alpha_2} p(\alpha) \,\dd\alpha,
\end{equation}
where $\alpha_1 = 2\eta_1$ and $\alpha_2 = 2\eta_2$.
Whence, $p$ is indeed a PDF with $\supp(p) = [\alpha_1, \alpha_2] \subseteq (-\pi,\pi)$, and we can safely insert it in Eq.~\eqref{eq:refPoint} to get
\begin{equation}\label{eq:microfacetScatteringEq}
	L_o(\th_o) \cos(\th_o) = \int_{-\pi/2}^{\pi/2} p(\th_i+\th_o) L_i(\th_i) \cos(\th_i) \, \dd\th_i,
\end{equation}
\noindent which is the so-called \textit{microfacet scattering equation} for microfacets with orientations described by the probability density function $p$, via $p(2\eta) = \tilde{p}(\eta)/2$.

\subsection{Freeform Reflector Design}
Thus far, we have derived the scattered \textit{radiance} [W $\cdot$ rad$^{-1}$ $\cdot$ m$^{-1}$] from a single point --- Eq.~\eqref{eq:microfacetScatteringEq}.
Let us now consider what occurs when the point is located on a curve that traces a reflector.
We shall show how this situation yields a convolution integral for the outgoing \textit{intensity} [W $\cdot$ rad$^{-1}$], which can be used to design reflectors via an inverse specular problem by using deconvolution.
We will then derive the relations needed to compute the reflectors for both parallel and point sources.
More details about two-dimensional specular freeform reflector design can be found in Maes's thesis \cite[Chs.~$1-6$]{maesMathematicalMethods2D1994}.

\subsubsection{Intensities}\label{sec:convolution}
Let $s \in [0, \ell]$, parametrising a curve of length $\ell$, be the arc length.
The point-wise microfacet scattering equation, Eq.~\eqref{eq:microfacetScatteringEq}, then becomes
\begin{equation}
	L_o(s,\th_o) \cos(\th_o) = \int_{-\pi/2}^{\pi/2}  p(\th_i+\th_o) L_i(s,\th_i) \cos(\th_i)	\, \dd\th_i,
\end{equation}
where $\th_o$ and $\th_i$ are measured from the unit normal $\un$ of the reflector at $s$, and $L_i(s,\th_i)$, $L_o(s,\th_o)$ are the incident and outgoing radiances at $s$, respectively.
Note that we have assumed that the scattering is independent of the location along the reflector since $p$ does not depend on $s$, i.e., the surface is assumed to be isotropic.

For a perfectly smooth reflector, an incident ray is specularly reflected from direction $\th_i$ into direction $\th_r = -\th_i$.
Defining a \textit{virtual specularly reflected radiance} $L_r(s,\th_r) := L_i(s,-\th_r)$, allows us to write the above relation as
\begin{equation}\label{eq:refPoint1}
	L_o(s,\th_o) \cos(\th_o) = \int_{-\pi/2}^{\pi/2}  p(\th_o-\th_r) L_r(s,\th_r) \cos(\th_r) \, \dd\th_r.
\end{equation}
This convolution integral allows the interpretation that an incident ray is first reflected into the specular direction $\ut$ at $s$ and then scattered into an off-specular direction $\uu$ due to a perturbation of the normal at $s$ described probabilistically using $p$.
This interpretation was explored further in our previous work \cite{kronbergModellingSurfaceLight2021a}.

\begin{figure}[H]
	\centering
	\includegraphics[width=0.5\linewidth,page=1]{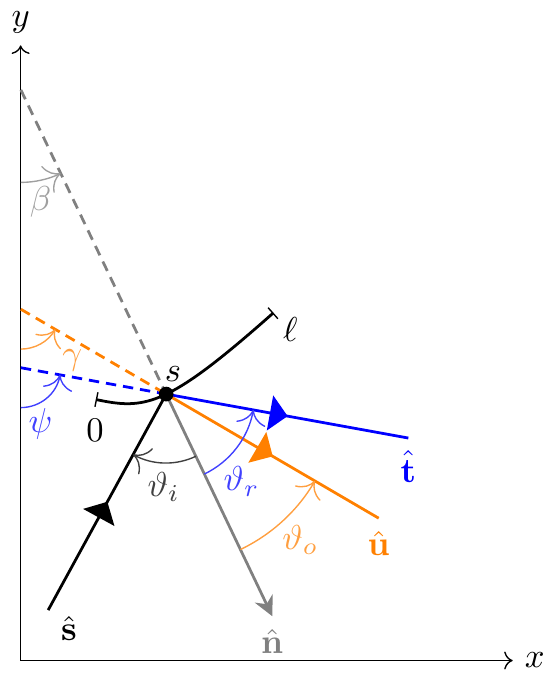}
	\captionsetup{width=\linewidth}
	\caption{The relation between the local and global angles; $\th_i < 0$ and $\beta$, $\gamma$, $\psi$, $\th_r$, $\th_o > 0$.}
	\label{fig:reflector_2}
\end{figure}

We now introduce so-called global angles.
By this, we mean that they are measured with respect to a static coordinate system rather than the local unit normal $\un$ --- see Fig.~\ref{fig:reflector_2}.
In particular, let $\beta$ be the direction of $\un$ at $s$, let $\psi$ be the specular direction, and let $\gamma$ be the scattered direction in the global coordinate system, i.e.,
\begin{equation}
	\un = \big(\sin(\beta),-\cos(\beta)\big)^\intercal, \quad \ut = \big(\sin(\psi),-\cos(\psi)\big)^\intercal, \quad \uu = \big(\sin(\gamma),-\cos(\gamma)\big)^\intercal.
\end{equation}
By inspecting Fig.~\ref{fig:reflector_2}, we see that (after some trigonometry)
\begin{equation}
	\th_r = \psi - \beta \qq{and} \th_o = \gamma - \beta.
\end{equation}

\noindent Transforming $\th_o$ to $\gamma$ and $\th_r$ to $\psi$ in Eq.~\eqref{eq:refPoint1} yields
\begin{equation}
	L_o(s,\gamma - \beta) \cos(\gamma - \beta) = \int_{-\pi}^{\pi}  p(\gamma - \psi) L_r(s,\psi - \beta) \cos(\psi - \beta) \, \dd\psi,
\end{equation}
where we have used the fact that $p$ has finite support $[\eta_1,\eta_2] \subseteq (-\pi/2,\pi/2)$ to extend the integration limits to the full circle.

Suppose that we are interested in \textit{all} light scattered by the surface in some global direction $\gamma$.
In this case, we must integrate over the whole reflector, keeping in mind that the normal $\un$, and hence $\beta$, changes as we traverse the reflector.
That is,
\begin{equation}
	\int_0^\ell L_o\big(s,\gamma - \beta(s)\big) \cos\big(\gamma - \beta(s)\big) \, \dd s = \int_0^\ell \int_{-\pi}^{\pi}  p(\gamma - \psi) L_r\big(s,\psi - \beta(s)\big) \cos\big(\psi - \beta(s)\big) \, \dd\psi \, \dd s.
\end{equation}
Notice that the left-hand side only depends on $\gamma$, so that we can define it as $h(\gamma)$ [W $\cdot$ rad$^{-1}$].
It represents the \textit{scattered \textbf{intensity}} distribution in direction $\gamma$.
By changing the order of integration on the right-hand side, we get
\begin{equation}
	h(\gamma) = \int_{-\pi}^{\pi} p(\gamma - \psi) \underbrace{\int_0^\ell L_r\big(s,\psi - \beta(s)\big) \cos\big(\psi - \beta(s)\big) \, \dd s}_{=: g(\psi)} \dd\psi,
\end{equation}
where $g(\psi)$ also has units [W $\cdot$ rad$^{-1}$] and represents a \textit{virtual specular \textbf{intensity}} distribution in direction $\psi$.
In total, we thus have that the outgoing intensity $h$ in some direction $\gamma$ is given by
\begin{equation}\label{eq:reflectorScatteringEq}
	h(\gamma) = \int_{-\pi}^{\pi} p(\gamma - \psi) \, g(\psi) \, \dd \psi.
\end{equation}
This is a convolution integral, which is often denoted as an asterisk operator; Eq.~\eqref{eq:reflectorScatteringEq} thus has the shorthand notation $h(\gamma) = (p * g)(\gamma)$.
We will occasionally use this notation, or simply $h = p * g$, referring to the convolution integral in Eq.~\eqref{eq:reflectorScatteringEq}.
Note that this is a general relation in that it is independent of the light source, which is intuitive since the scattering event occurs at the reflector surface, independently of how the light arrives there.

We shall discuss in detail how this model can be used to design specular reflectors whilst taking surface scattering into account in Sec.~\ref{sec:numerical}.
However, the key is to prescribe the desired target distribution $h$ and the PDF $p$ describing the orientations of the microfacets.
Then, a simple deconvolution procedure can be used to obtain the virtual specular intensity distribution $g$, which is then utilised to compute the reflector surface using well-established specular design procedures.
We note that due to surface scattering, the $g$ we computed by deconvolution is \textit{not} achieved in practice by the reflector manufactured using the scattering surface.
Instead, the deconvolved $g$ is merely used as a tool to utilise well-established specular reflector design methods for designing reflectors with surfaces exhibiting scattering.
Of course, a perfect mirror in the shape of the reflector computed using the deconvolved specular distribution \textit{would} achieve this $g$.

\subsubsection{Parallel Sources}\label{sec:specParallel}
Before discussing the practical considerations of using this theory for reflector design by showing some numerical examples, let us briefly derive the relations needed for specular reflector design, starting with parallel sources.

Recall Fig.~\ref{fig:sources} and definition \ref{def:parallelSource}, where the parallel source was first introduced.
In particular, consider a parallel source with $\th_s = 0$ so that the rays travel along the positive $y$-axis.
Furthermore, let $x$ be the coordinate running along the source, and let $\ell_s$ be the length of the source.
The reflector is parametrised by $\mathbf{r}(x) = \big(x, u(x)\big)^\intercal$, where $u(x) > 0$ for all $x \in [0,\ell_s]$ is a smooth height function.
Thus, a unit tangent vector $\hat{\boldsymbol{\tau}}$ to the reflector is by definition given by $\hat{\boldsymbol{\tau}} = \mathbf{r}'(x)/\abs{\mathbf{r}'(x)} = \big(1, u'(x)\big)^\intercal/\sqrt{1+u'(x)^2}$, and the unit normal $\un$ pointing towards the light source is given by
\begin{equation}
	\un = \mathbf{R}(-\pi/2) \hat{\boldsymbol{\tau}} = \frac{1}{\sqrt{1+u'(x)^2}} \, \big(u'(x), -1\big)^\intercal,
\end{equation}
where the rotation matrix $\mathbf{R}$ was introduced in Eq.~\eqref{eq:rotMatrix}.
Recall that all source rays $\us$ are given by $\us \equiv \ue_y = (0, 1)^\intercal$, so that
\begin{equation}\label{eq:sn_parallel}
	\us \cdot \un = \frac{-1}{\sqrt{1+u'(x)^2}}.
\end{equation}
The vectorial law of reflection may be expressed as $\ut = \us - 2(\us \cdot \un)\un$, so that
\begin{equation}\label{eq:st}
	\us \cdot \ut = 1 - 2(\us \cdot \un)^2.
\end{equation}

\begin{figure}[ht!]
	\centering
	\includegraphics[width=0.7\linewidth,page=2]{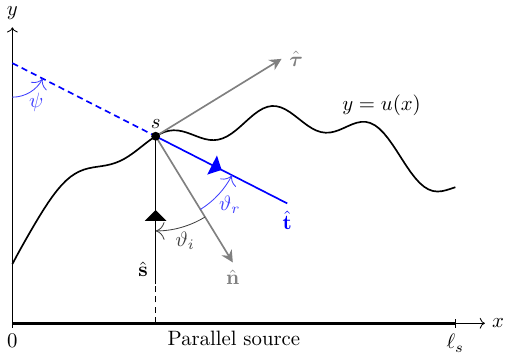}
	\captionsetup{width=0.85\linewidth}
	\caption{Parallel illumination of a reflector traced by the curve $y = u(x)$; the relevant unit vectors and angles close to the intersection point $s$ are shown; $\th_i < 0$ and $\psi$, $\th_r > 0$.}
	\label{fig:parallelRefl}
\end{figure}

Now, from the geometry near the intersection point $s$ between a source ray and the reflector surface (Fig.~\ref{fig:parallelRefl}), we get
\begin{equation}
	2 \th_r = \psi.
\end{equation}

\noindent Then, by the definition of the dot product,
\begin{equation}
	- \us \cdot \ut = \cos(2\th_r) = \cos(\psi).
\end{equation}
Returning to Eq.~\eqref{eq:st}, we get
\begin{equation}
	-\cos(\psi) = 1 - 2(\us \cdot \un)^2.
\end{equation}
Finally, using Eq.~\eqref{eq:sn_parallel}, we find that
\begin{equation}
	-\cos(\psi) = 1 - 2\,\bigg(\frac{1}{1+u'(x)^2}\bigg).
\end{equation}
Solving for $u'(x)$ gives the ordinary differential equations (ODEs)
\begin{equation}
	u'(x) = \pm \sqrt{\frac{1-\cos(\psi)}{1+\cos(\psi)}}.
\end{equation}
The sign of the root is related to the direction of the reflected ray.
In particular, using the vectorial law of reflection, $\ut = \us - 2(\us \cdot \un)\un$, we get $\ut = (\pm \sin(\psi), -\cos(\psi))^\intercal$ depending on the sign of the root.
Since $\ut = (\sin(\psi), -\cos(\psi))^\intercal$, we use the positive root together with the tangent half-angle relation \cite[p.~127]{radeMathematicsHandbookScience2011} to get the appropriate ODE
\begin{equation}
	u'(x) = \tan(\frac{\psi}{2}).
\end{equation}
To obtain the height function $u(x)$, we must solve this ODE with some initial value for the distance of the first point on the reflector from the source, $u(0) = u_0$, making it an initial value problem (IVP).
Before discussing how to solve this IVP, note that $\psi$ is a map from a position on the source $x$ onto a direction $\psi$.
Denoting this map $m(x)$, i.e., $\psi = m(x)$, we get the following IVP
\begin{equation}\label{eq:uIVP_parallel}
	\begin{cases}
		\mathlarger{u'(x) = \tan(\frac{m(x)}{2})}, \quad 0 < x \leq \ell_s;\\
		\mathlarger{u(0) = u_0}.
	\end{cases}
\end{equation}
Note that a representation of the solution to this IVP is
\begin{equation}\label{eq:IVPSolPar}
	u(x) = u_0 + \int_0^x \tan(\frac{m(\tilde{x})}{2}) \, \dd\tilde{x}.
\end{equation}

\paragraph{Optical map}
\noindent We will now show how to construct two types of solutions to the above IVP.
In particular, suppose the optical map $m(x) = m_+(x)$ is monotonically increasing on $x \in [0,\ell_s]$.
Let $\forall x \in [0,\ell_s]$: $M(x) > 0$ [W $\cdot$ m$^{-1}$] be the parallel source exitance, and recall that $\forall \psi \in [\psi_1, \psi_2] \subseteq (-\pi/2,\pi/2)$: $g(\psi) > 0$ [W $\cdot$ rad$^{-1}$] is the specular intensity.
Since $m_+(x)$ is monotonically increasing, conservation of flux gives
\begin{equation}
	\forall x \in [0,\ell_s]\!: \int_{0}^x M(\tilde{x}) \, \dd \tilde{x} = \int_{\psi_1}^{m_+(x)} g(\psi) \, \dd\psi,
\end{equation}
which can be differentiated with respect to $x$ to give the equivalent initial value problem:
\begin{equation}\label{eq:opticalMap_parallelPlus}
	\begin{cases}
		\mathlarger{m_+'(x) = \frac{M(x)}{g\big(m_+(x)\big)}}, \quad 0 < x \leq \ell_s;\\
		\mathlarger{m_+(0) = \psi_1}.
	\end{cases}
\end{equation}
On the other hand, suppose the optical map $m(x) = m_-(x)$ is monotonically decreasing on $x \in [0,\ell_s]$, then conservation of flux gives
\begin{equation}
	\forall x \in [0,\ell_s]\!: \int_{0}^x M(\tilde{x}) \, \dd \tilde{x} = \int_{m_-(x)}^{\psi_2} g(\psi) \, \dd\psi,
\end{equation}
which can be differentiated with respect to $x$ to give the equivalent IVP:
\begin{equation}\label{eq:opticalMap_parallelMinus}
	\begin{cases}
		\mathlarger{m_-'(x) = -\frac{M(x)}{g\big(m_-(x)\big)}}, \quad 0 < x \leq \ell_s;\\
		\mathlarger{m_-(0) = \psi_2}.
	\end{cases}
\end{equation}

\subsubsection{Point sources}\label{sec:specPoint}
\begin{figure}[H]
	\centering
	\includegraphics[width=0.65\linewidth,page=3]{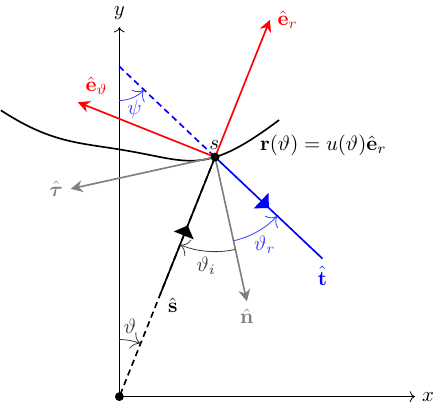}
	\captionsetup{width=0.9\linewidth}
	\caption{Point source illumination of a reflector traced by the curve $\mathbf{r}(\th)=u(\th)\ue_r$; the relevant unit vectors and angles close to the intersection point $s$ are shown; $\th$, $\th_i < 0$ and $\psi$, $\th_r > 0$.}
	\label{fig:pointRefl}
\end{figure}

We now move on to point sources; recall Fig.~\ref{fig:sources} and definition \ref{def:pointSource}, where they were first introduced.
In particular, consider a point source with $x_s = 0$, i.e., centred at the origin of an $xy$-coordinate system.
Let $\th$ be the angle between the positive $y$-axis and a source ray --- see Fig.~\ref{fig:pointRefl}.
Let $[\th_1,\th_2] \subseteq (-\pi/2,\pi/2)$ be the interval where the reflector, described using a smooth, continuous radius function $u(\th) > 0$, exists.
The reflector is thus parametrised by $\mathbf{r}(\th) = u(\th) \ue_r$, where $\ue_r = \big(\!-\sin(\th),\cos(\th)\big)^\intercal$ is the radial unit vector in our polar coordinate system.
Thus, a unit tangent vector $\hat{\boldsymbol{\tau}}$ to $\mathbf{r}(\th)$ is given by $\hat{\boldsymbol{\tau}} = \mathbf{r}'(\th)/\abs{\mathbf{r}'(\th)} = (u'(\th)\ue_r + u(\th)\ue_\th)/\sqrt{u(\th)^2 + u'(\th)^2}$, where $\ue_\th = \ue_r' = \big(\!-\cos(\th),-\sin(\th)\big)^\intercal$ is the angular unit vector in our polar coordinate system.
Whence, the unit normal $\un$ pointing towards the light source is given by
\begin{equation}
	\un = \mathbf{R}(\pi/2) \hat{\boldsymbol{\tau}} = \frac{-u(\th)\ue_r + u'(\th)\ue_\th}{\sqrt{u(\th)^2 + u'(\th)^2}}.
\end{equation}
Introducing $v(\th) := \ln\!\big(u(\th)\big)$, we get
\begin{equation}
	\un = \frac{- \ue_r + v'(\th)\ue_\th}{\sqrt{1 + v'(\th)^2}}.
\end{equation}
Note that all source rays $\us$ are given by $\us \equiv \ue_r$, so that
\begin{equation}\label{eq:sn_point}
	\us \cdot \un = -\frac{1}{\sqrt{1 + v'(\th)^2}}.
\end{equation}

From the geometry near the intersection of a source ray and the reflector surface (Fig.~\ref{fig:pointRefl}) and the definition of the dot product, we get, after some trigonometry (note that $\psi - \th = \th_r - \th_i$ and recall that $\th_i = -\th_r$),
\begin{equation}
	- \us \cdot \ut = \cos(\th - \psi).
\end{equation}

\noindent Since the law of reflection, $\ut = \us - 2(\us \cdot \un)\un$, is valid for point sources as well, so is Eq.~\eqref{eq:st}, and we get
\begin{equation}
	-\cos(\th - \psi) = 1 - 2(\us \cdot \un)^2.
\end{equation}
Finally, using Eq.~\eqref{eq:sn_point}, we find that
\begin{equation}
	-\cos(\th - \psi) = 1 - 2\,\bigg(\frac{1}{1 + v'(\th)^2}\bigg).
\end{equation}

Solving for $v'(x)$ gives the ODEs
\begin{equation}
	v'(\th) = \pm \sqrt{\frac{1-\cos(\th-\psi)}{1+\cos(\th-\psi)}}.
\end{equation}
Note that this time, $\psi$ is a map from the angle of a source ray, $\th$, onto a specular direction $\psi$, i.e., $\psi = m(\th)$.
Comparing $\ut = (\sin(\psi), -\cos(\psi))^\intercal$ with $\ut = \us - 2(\us \cdot \un) \un = -\cos(\th - \psi) \ue_r \pm \sin(\th - \psi) \ue_\th$, we conclude that we must once again use the positive root.
By using the tangent half-angle relation \cite[p.~127]{radeMathematicsHandbookScience2011}, we thus get
\begin{equation}
	v'(\th) = \tan(\frac{\th - m(\th)}{2}).
\end{equation}
Using the initial condition $u(\th_1) = u_0$, we get the IVP
\begin{equation}\label{eq:uIVP_point}
	\begin{cases}
		\mathlarger{v'(\th) = \tan(\frac{\th - m(\th)}{2})}, \quad \th_1 < \th \leq \th_2;\\
		\mathlarger{v(\th_1) = \ln(u_0)},
	\end{cases}
\end{equation}
with solutions given by
\begin{equation}\label{eq:IVPSolPoint}
	v(\th) = v(\th_1) + \int_{\th_1}^\th \tan(\frac{\tilde{\th} - m(\tilde{\th})}{2}) \, \dd\tilde{\th}.
\end{equation}
After solving the IVP for $v(\th)$, we obtain the radius function $u(\th) = \exp\!\big(v(\th)\big)$, which fully determines the reflector.

\paragraph{Optical map}
\noindent Similar to the parallel source case, we will now show how to construct two types of solutions to the above IVP.
Suppose the optical map $m(\th) = m_+(\th)$ is monotonically increasing on $\th \in [\th_1,\th_2] \subseteq (-\pi/2,\pi/2)$.
Let $\forall \th \in [\th_1,\th_2]$: $f(\th) > 0$ [W $\cdot$ rad$^{-1}$] be the point source intensity, and recall that $\forall \psi \in [\psi_1, \psi_2] \subseteq (-\pi/2,\pi/2)$: $g(\psi) > 0$ [W $\cdot$ rad$^{-1}$] is the specular intensity.
Since $m_+(\th)$ is monotonically increasing, conservation of flux gives
\begin{equation}
	\forall \th \in [\th_1,\th_2]\!: \int_{\th_1}^\th f(\tilde{\th}) \, \dd \tilde{\th} = \int_{\psi_1}^{m_+(\th)} g(\psi) \, \dd\psi,
\end{equation}
which can be differentiated with respect to $\th$ to give the equivalent IVP:
\begin{equation}\label{eq:opticalMap_pointPlus}
	\begin{cases}
		\mathlarger{m_+'(\th) = \frac{f(\th)}{g\big(m_+(\th)\big)}}, \quad \th_1 < \th \leq \th_2;\\
		\mathlarger{m_+(\th_1) = \psi_1}.
	\end{cases}
\end{equation}
On the other hand, suppose the optical map $m(\th) = m_-(\th)$ is monotonically decreasing on $\th \in [\th_1,\th_2]$, then conservation of flux gives
\begin{equation}
	\forall \th \in [\th_1,\th_2]\!: \int_{\th_1}^\th f(\tilde{\th}) \, \dd \tilde{\th} = \int_{m_-(\th)}^{\psi_2} g(\psi) \, \dd\psi,
\end{equation}
which can be differentiated with respect to $\th$ to give the equivalent IVP:
\begin{equation}\label{eq:opticalMap_pointMinus}
	\begin{cases}
		\mathlarger{m_-'(\th) = -\frac{f(\th)}{g\big(m_-(\th)\big)}}, \quad \th_1 < \th \leq \th_2;\\
		\mathlarger{m_-(\th_1) = \psi_2}.
	\end{cases}
\end{equation}

\section{Numerical Examples}\label{sec:numerical}
This section shows how to use the equations we just derived.
In particular, we shall discuss the proposed workflow of an optical engineer and show how to solve the relevant differential equations for a selection of examples.
For more examples and further discussion, please see our previous work \cite{kronbergModellingSurfaceLight2021a}.

\subsection{Verification}\label{sec:validation}
We have written a custom raytracer to verify our model, which implements the microfacets introduced when deriving the scattering model.
First, the reflector surface was numerically computed on a fixed grid (either on $x$ or $\th$, depending on the source type) by solving the IVPs for $m$ and $u$.
Then, the raytracing starts with rays sampled from the desired source distribution $M(x)$ or $f(\th)$.
A sampled ray strikes the reflector at some point $s$, where the normal $\un$ is determined using linear interpolation between known normals (i.e., at the fixed grid points where the reflector surface and normals were computed) using Matlab's \texttt{interp1} routine. The law of reflection, $\ut = \us - 2(\us\cdot\un)\un$, is subsequently applied to get the specular ray at $s$.
Next, the scattered ray $\uu$ is computed by applying a rotation $\mathbf{R}(\eta)$ with $\eta \in [\eta_1,\eta_2] \subseteq (-\pi/2,\pi/2)$ sampled from the PDF $p(\eta)$, to $\un$ to get $\un^m = \mathbf{R}(\eta) \un$, and then the law of reflection, $\uu = \us - 2(\us\cdot\un^m)\un^m$, is applied.

The spatial or angular domains are divided into $N$ equispaced regions, so-called `bins', i.e., for some positive integer $j < N$: $[x_{j-1},x_j]$ or $[\th_{j-1},\th_j]$.
The bins of the source, specular and scattered rays are identified using Matlab's \texttt{dsearchn} nearest point search, and finally, the ray counts are converted to an exitance $E$ or intensity $I$ using
\begin{equation}\label{eq:RT_M}
	\forall j \in [1,N]\,: E_j = \frac{\text{Pr}(x_{j-1} \leq x < x_j)}{\Delta x} \int_{0}^{\ell_s} M(x) \, \dd x,
\end{equation}
or
\begin{equation}\label{eq:RT_I-point}
	\forall j \in [1,N]\,: I_j = \frac{\text{Pr}(\th_{j-1} \leq \th < \th_j)}{\Delta \th} \int_{\th_1}^{\th_2} f(\th) \, \dd \th,
\end{equation}
for illumination by parallel and point sources, respectively.
Here, the integral represents the total flux of the source, and $\text{Pr}(x_{j-1} \leq x < x_j)$ or $\text{Pr}(\th_{j-1} \leq \th < \th_j)$ is the number of rays in the $j$th bin divided by the total number of rays traced, i.e., the probability of falling in the $j$th bin, and $\Delta x$ or $\Delta \th$ is the size of the collection bins.

Naturally, Eq.~\eqref{eq:RT_M} is only used for the \textit{source rays} from a parallel source since the specular and scattered rays all result in an \textit{intensity}, which exists in angular space.
For the \textit{specular} and \textit{scattered rays} with a parallel source, the following expression is used:
\begin{equation}\label{eq:RT_I-parallel}
	\forall j \in [1,N]\,: I_j = \frac{\text{Pr}(\th_{j-1} \leq \th < \th_j)}{\Delta \th} \int_{0}^{\ell_s} M(x) \, \dd x.
\end{equation}
In other words, for a parallel source, $E$ approximates $M$ via Eq.~\eqref{eq:RT_M}, and $I$ approximates $g$ and $h$ via Eq.~\eqref{eq:RT_I-parallel}, whilst for a point source, $I$ approximates $f$, $g$ and $h$ via Eq.~\eqref{eq:RT_I-point}.
More details are available in \cite[p.~34]{filosaPhaseSpaceRay2018}.

To quantitatively evaluate the accuracy of our predicted scattered distributions, we shall use the root mean square (RMS) error given by
\begin{equation}\label{eq:h_RMS}
	\varepsilon(h,h^*) := \sqrt{\frac{1}{N} \sum_{n=1}^N \abs{h_n - h^*_n}^2}\,
\end{equation}
for $N$ collection bins of the raytraced scattered distribution, $h^*$, and the scattered distribution $h$.
We always denote raytraced distributions with an upper-index asterisk $(^*)$.
In addition, the \textbf{d}econvolved specular distribution is denoted $g_\mathrm{dc}$, whilst $h_\mathrm{rc} := g_\mathrm{dc} * p$ (recall Eq.~\eqref{eq:reflectorScatteringEq}) represents the ``\textbf{r}econvolved'' distribution.

\subsection{Parallel Source}\label{sec:examplesParallel}
We shall start with an example of reflector design using a parallel source.
The first example is shown in the box below, where
\begin{equation}
	\mathcal{N}(\theta; \mu,\sigma) := \frac{1}{\sigma\sqrt{2\pi}} \exp\Bigg({-}\frac{1}{2} \bigg(\frac{\theta - \mu}{\sigma}\bigg)^2\Bigg),
\end{equation}
represents the Gaussian distribution centred at $\mu$ with standard deviation $\sigma$.
The distributions $M$, $g$, and $p$ have finite support given by $[0,\ell_s]$, $[\psi_1, \psi_2]$, and $[\eta_1,\eta_2]$, respectively.
That is, we set the values to exactly zero outside these domains and renormalise as appropriate.
Since we first wish to verify our predicted scattered distribution $h$, we shall prescribe the known specular distribution $g$ and use this as our target function when designing the reflectors.
Thus, the verification step here is to check that the raytraced scattered distribution $h^*$ approaches the predicted scattered distribution $h$ from the convolution integral in Eq.~\eqref{eq:reflectorScatteringEq}.

\begin{mdframed}
	\textbf{Example \#1: Overlapping Gaussians}
	\begin{align*}
		&\text{$x$-range:} 							&[0, \ell_s] 				&= [0,1]\\
		&\text{$\psi$-range:} 						&[\psi_1, \psi_2] 			&= [-1.64,0.95]\\
		&\text{$\eta$-range:}						&[\eta_1,\eta_2]			&= [-\pi/2,\pi/2]\\
		&\text{Source distribution:} 				&M(x)						&= 1\\
		&\text{Specular target distribution:} 		&g(\psi) 					&= \mathcal{N}(\psi;-17^{\circ},20^{\circ}) + \mathcal{N}(\psi;23^{\circ},10^{\circ})\\
		&\text{Surface scattering function:} 		&p(\eta) 					&= \mathcal{N}(\eta; 0,10^{\circ})\\
		&\text{Scattered distribution prediction:}	&h(\gamma) 					&= (p*g)(\gamma)
	\end{align*}
\end{mdframed}

\noindent Since the specular target distribution $g(\psi)$ is a sum of two Gaussians, it does not have a finite support.
In order to clamp the support $[\psi_1, \psi_2] \subseteq (-\pi/2, \pi/2)$, we fixed $\epsilon = 10^{-3}$ and found the points where $g(\psi) = \epsilon$.
We then set the values of $g(\psi)$ to zero outside $[\psi_1, \psi_2]$ and renormalised it such that energy was conserved.
The Gaussian distribution $p(\eta)$ was given finite support $[\eta_1,\eta_2] = [-\pi/2,\pi/2]$ in a similar manner, and it was then renormalised to unity on this domain, as required for it to be a probability density function.
The predicted scattered distribution $h(\gamma)$ was computed using Matlab's \texttt{conv} function.

\begin{figure}[htbp]
	\centering
	\includegraphics[width=0.5\linewidth]{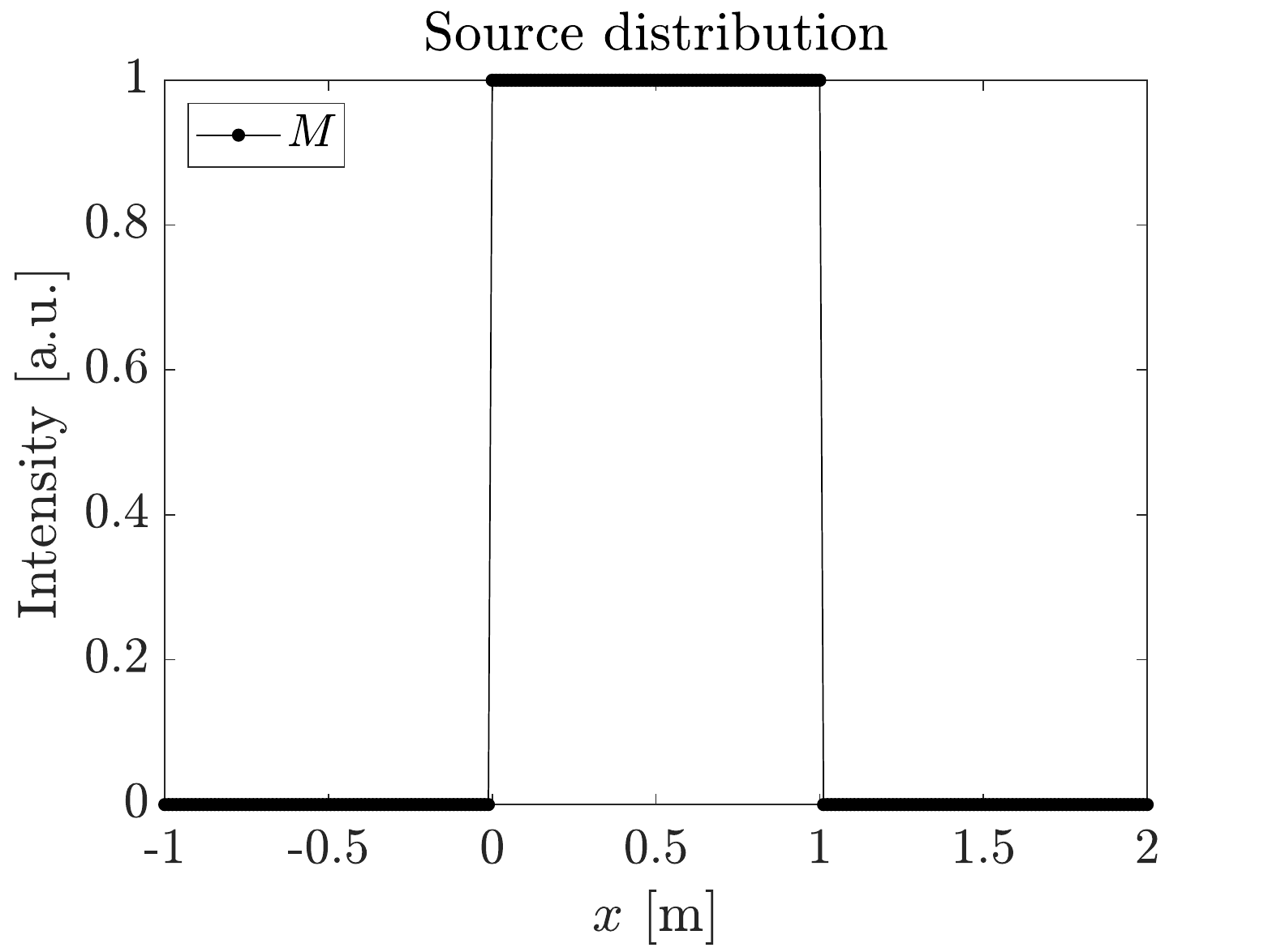}%
	\includegraphics[width=0.5\linewidth]{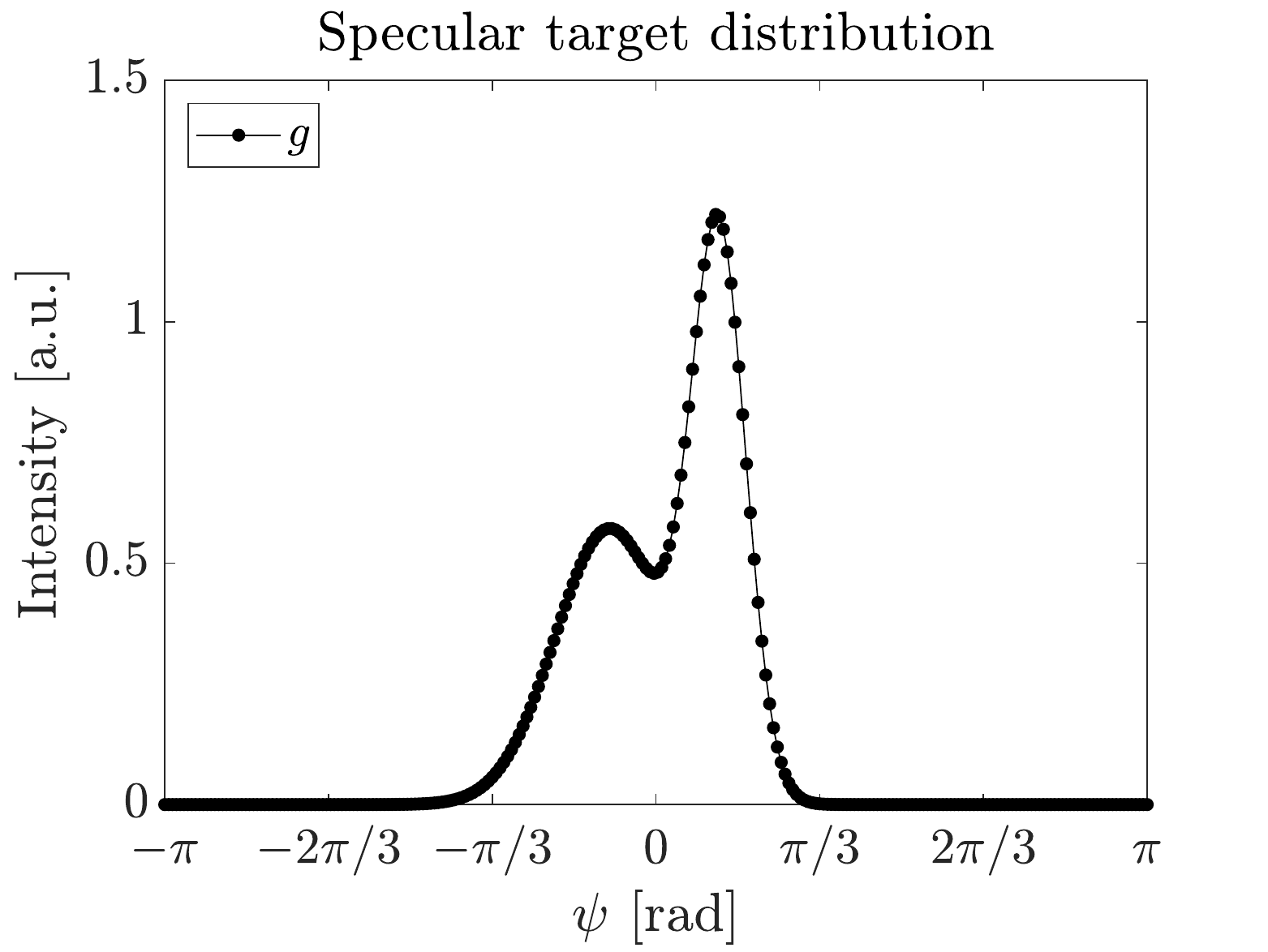}\\[5pt]
	\includegraphics[width=0.5\linewidth]{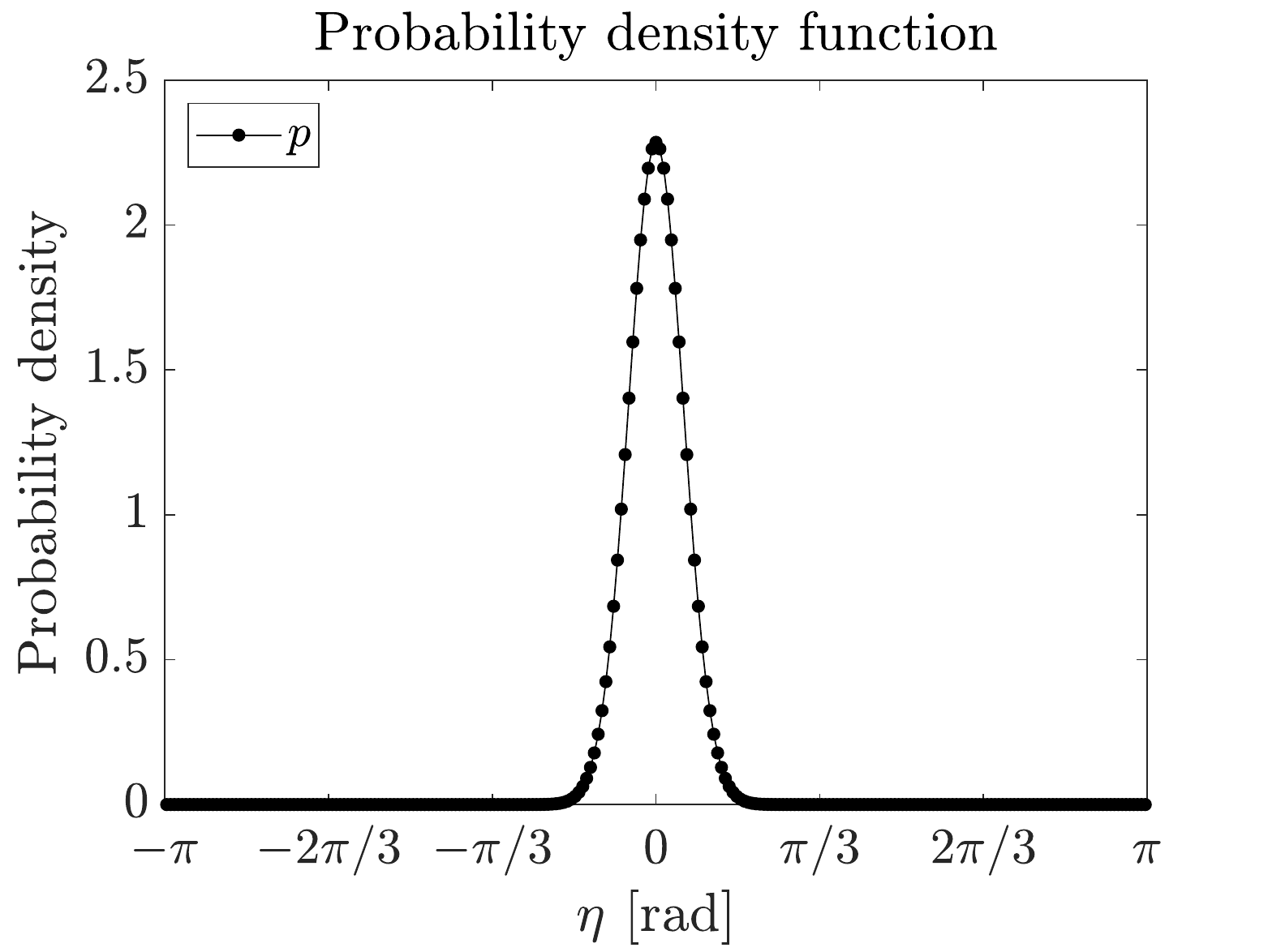}%
	\includegraphics[width=0.5\linewidth]{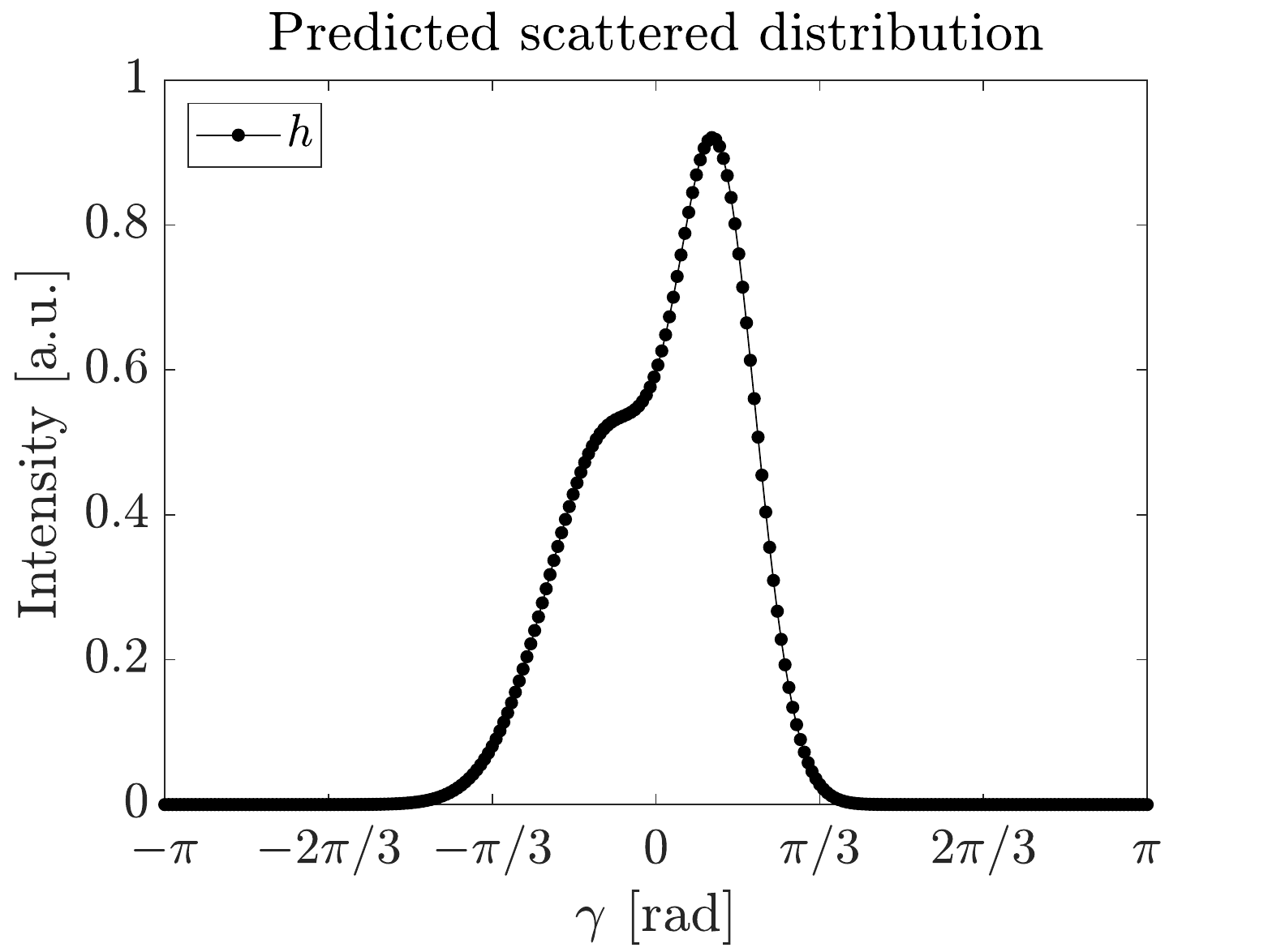}
	\captionsetup{width=\linewidth}
	\caption{Distributions in Example \#1; $256$ sample points.}
	\label{fig:example_1-1}
\end{figure}

The distributions in this example are shown in Fig.~\ref{fig:example_1-1}.
Note that all distributions are plotted on a range exceeding the previously mentioned domains.
This is called ``zero-padding'', which is necessary when performing convolution since the scattered distribution broadens so that it extends outside the support of $g$.
When numerically computing the reflectors, we have chosen $256$ sample points for our distributions.
The reason for using numerical methods here is that, whilst we do have exact expressions for all the distributions for this problem, we want the method to be as general as possible.

\begin{figure}[H]
	\centering
	\includegraphics[width=0.5\linewidth]{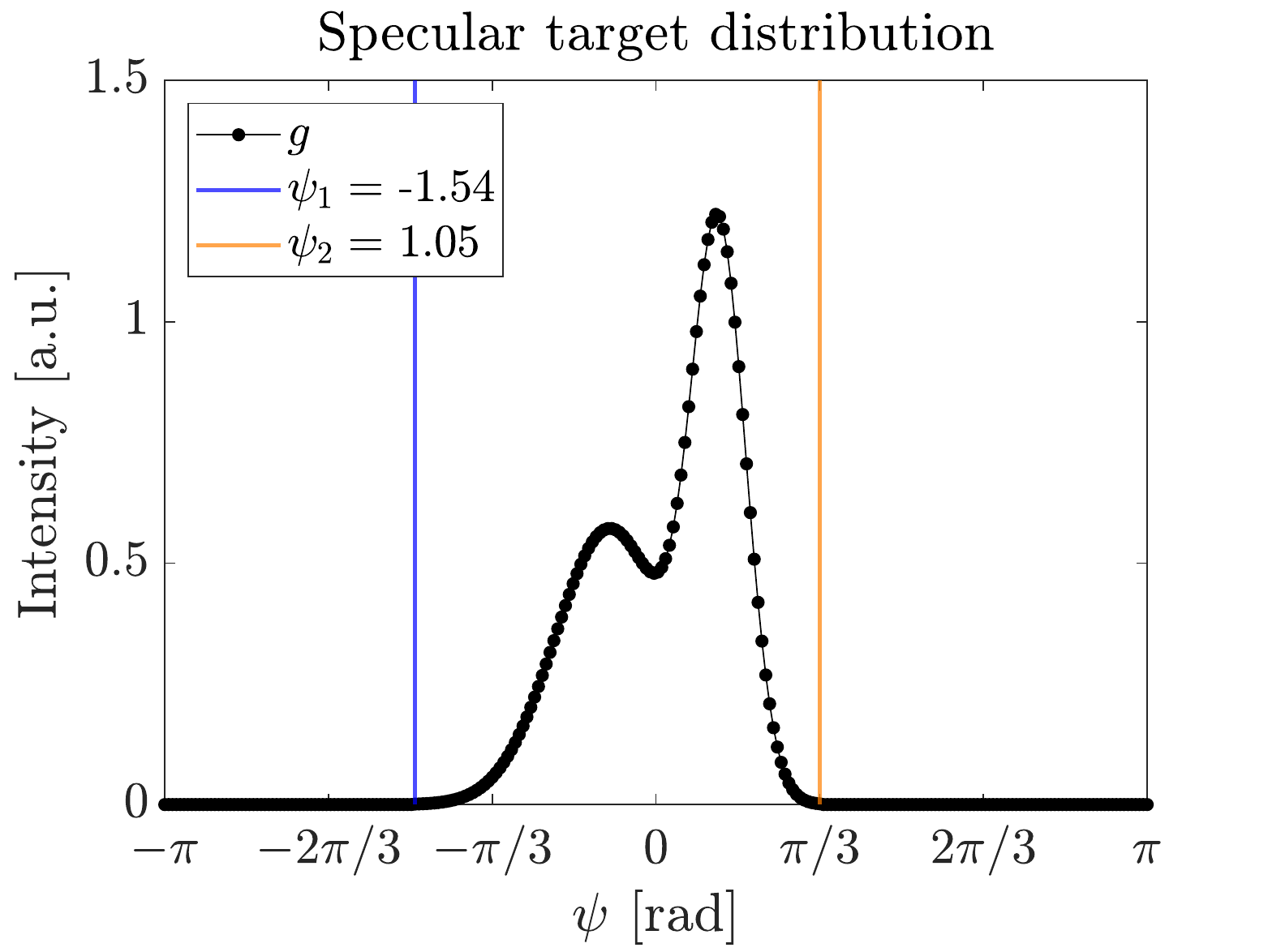}\\[5pt]
	\includegraphics[width=0.5\linewidth]{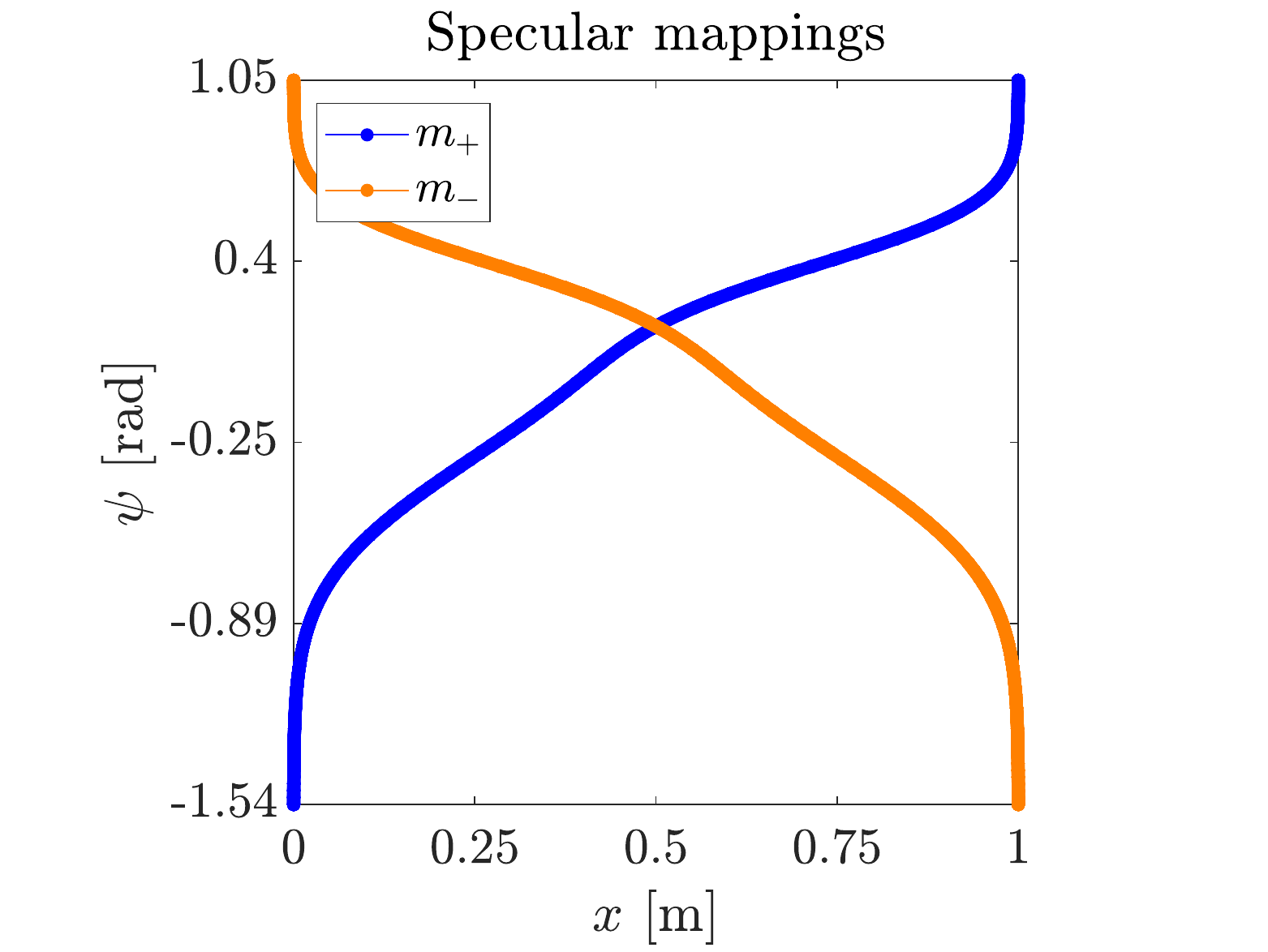}%
	\includegraphics[width=0.5\linewidth]{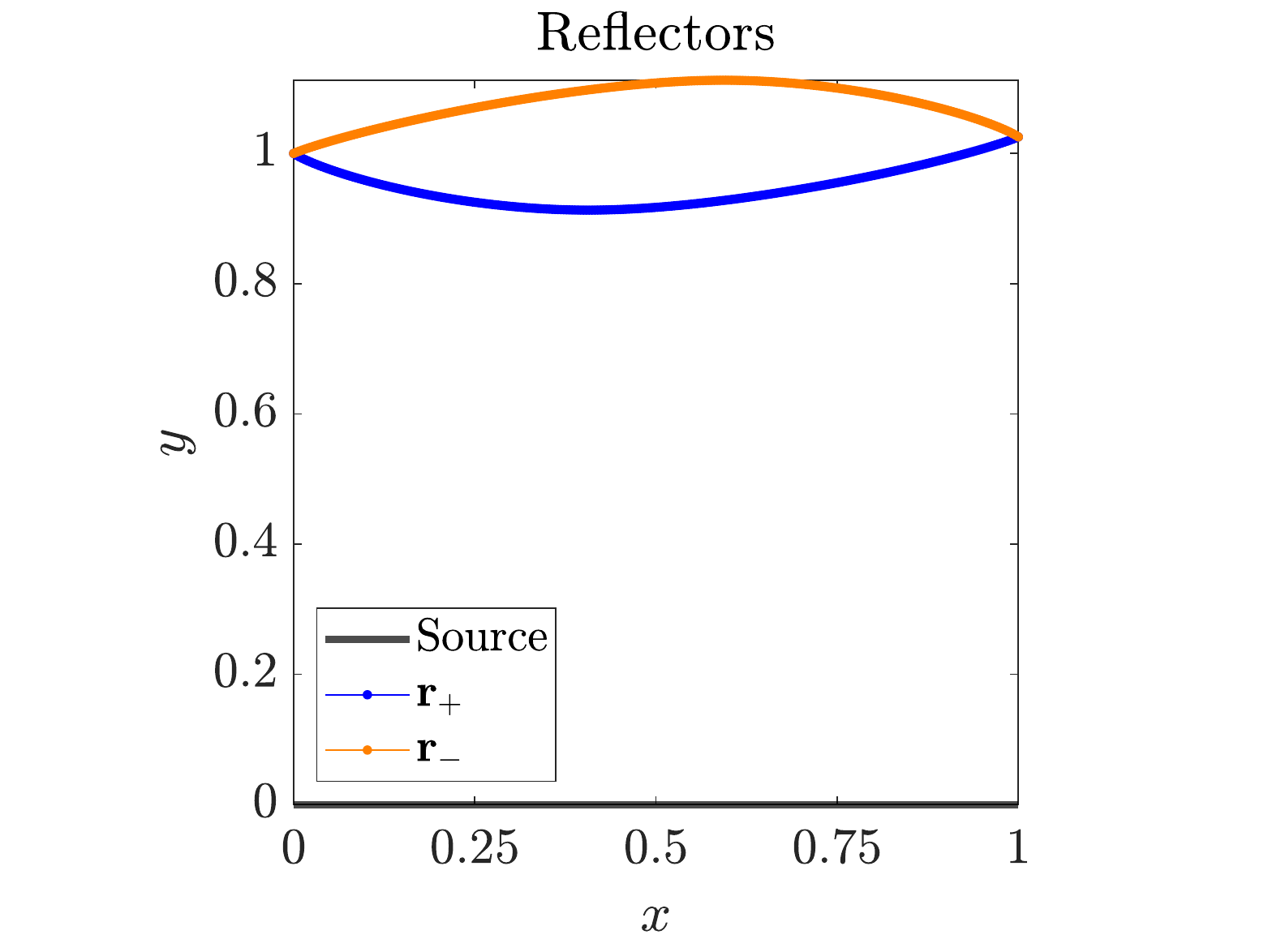}
	\captionsetup{width=\linewidth}
	\caption{The finite domain of $g$, specular mappings $m_\pm$ and associated reflectors $\mathbf{r}_\pm$ in Example \#1; 256 sample points of $g$ and $\mathbf{r}_\pm$.}
	\label{fig:example_1-2}
\end{figure}

To proceed, we now show the finite support $[\psi_1, \psi_2]$ of $g(\psi)$ (found using $g(\psi) = \epsilon$), the specular mappings $m_\pm$, and the computed reflectors in Fig.~\ref{fig:example_1-2}, for $256$ sample points of the distributions.
The specular mappings $m_\pm$ were computed on a variable grid using Matlab's \texttt{ode15s} solver with tolerances \texttt{RelTol} $=10^{-12}$ and \texttt{AbsTol} $=10^{-14}$, whilst the reflectors were computed using the integral representation of the IVP, i.e., Eq.~\eqref{eq:IVPSolPar}, using Matlab's \texttt{integral} command, on a fixed grid with 256 sample points.

\begin{figure}[H]
	\centering
	\includegraphics[width=0.5\linewidth]{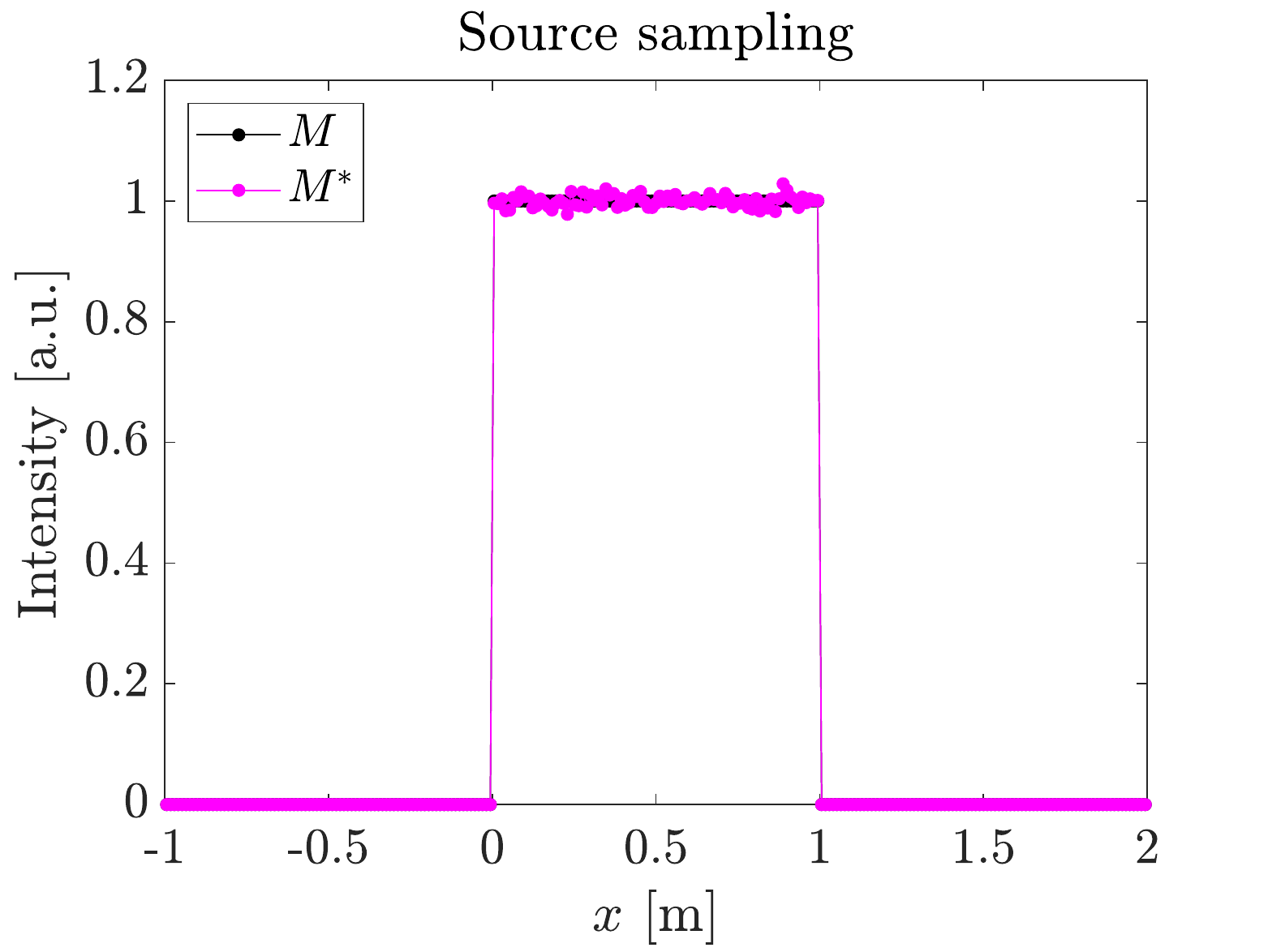}%
	\includegraphics[width=0.5\linewidth,page=1]{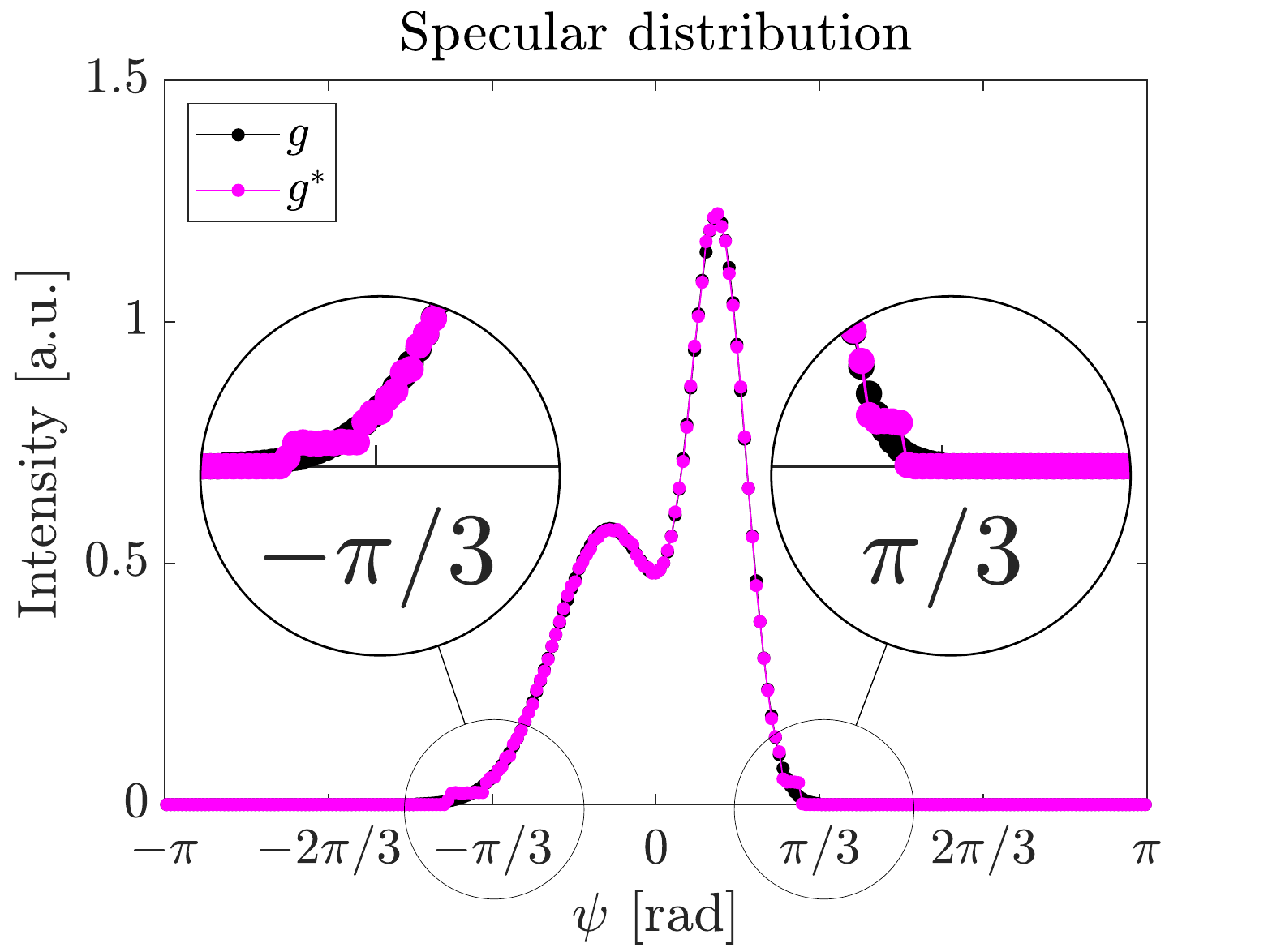}\\[5pt]
	\includegraphics[width=0.5\linewidth]{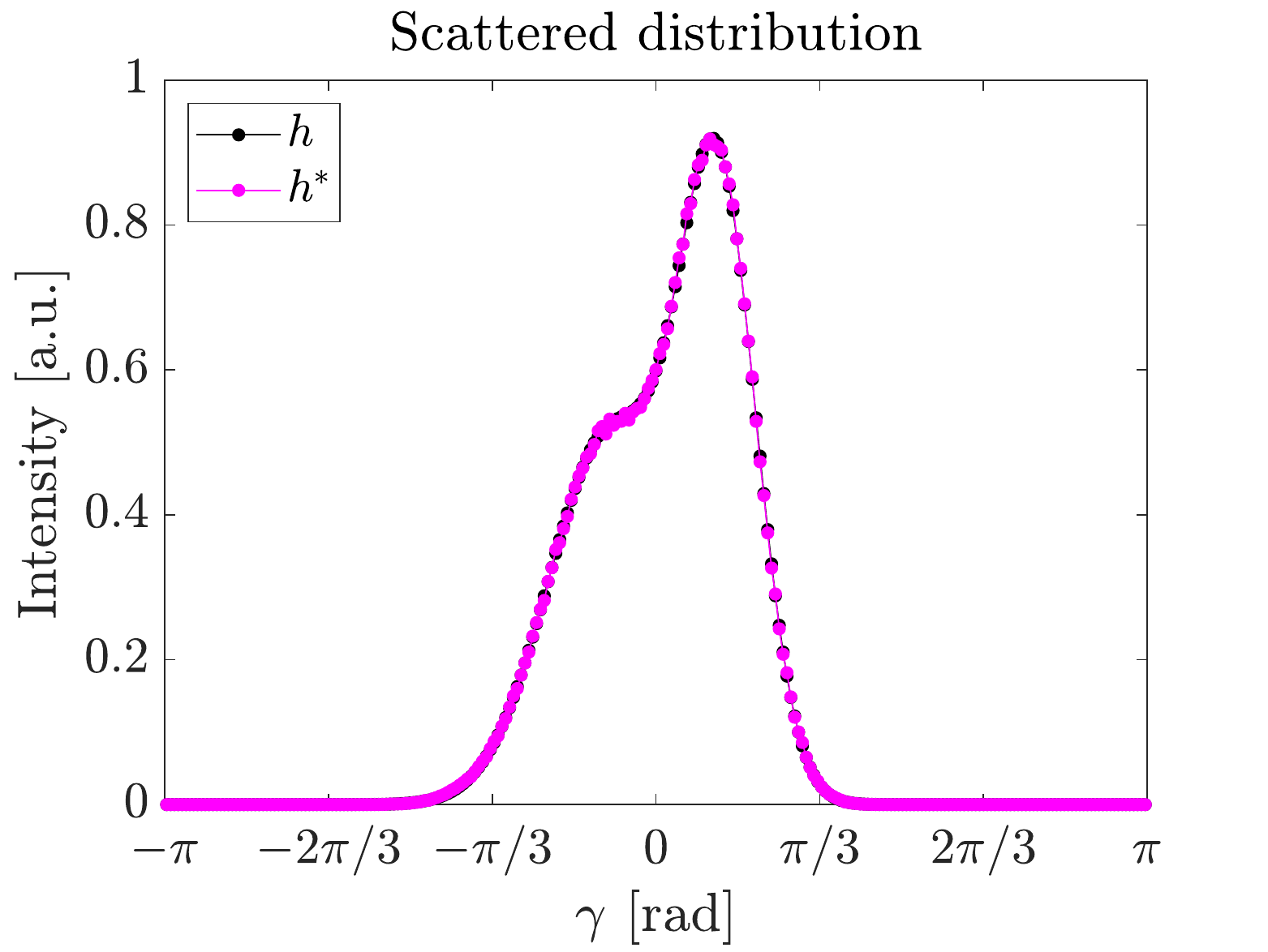}%
	\includegraphics[width=0.5\linewidth]{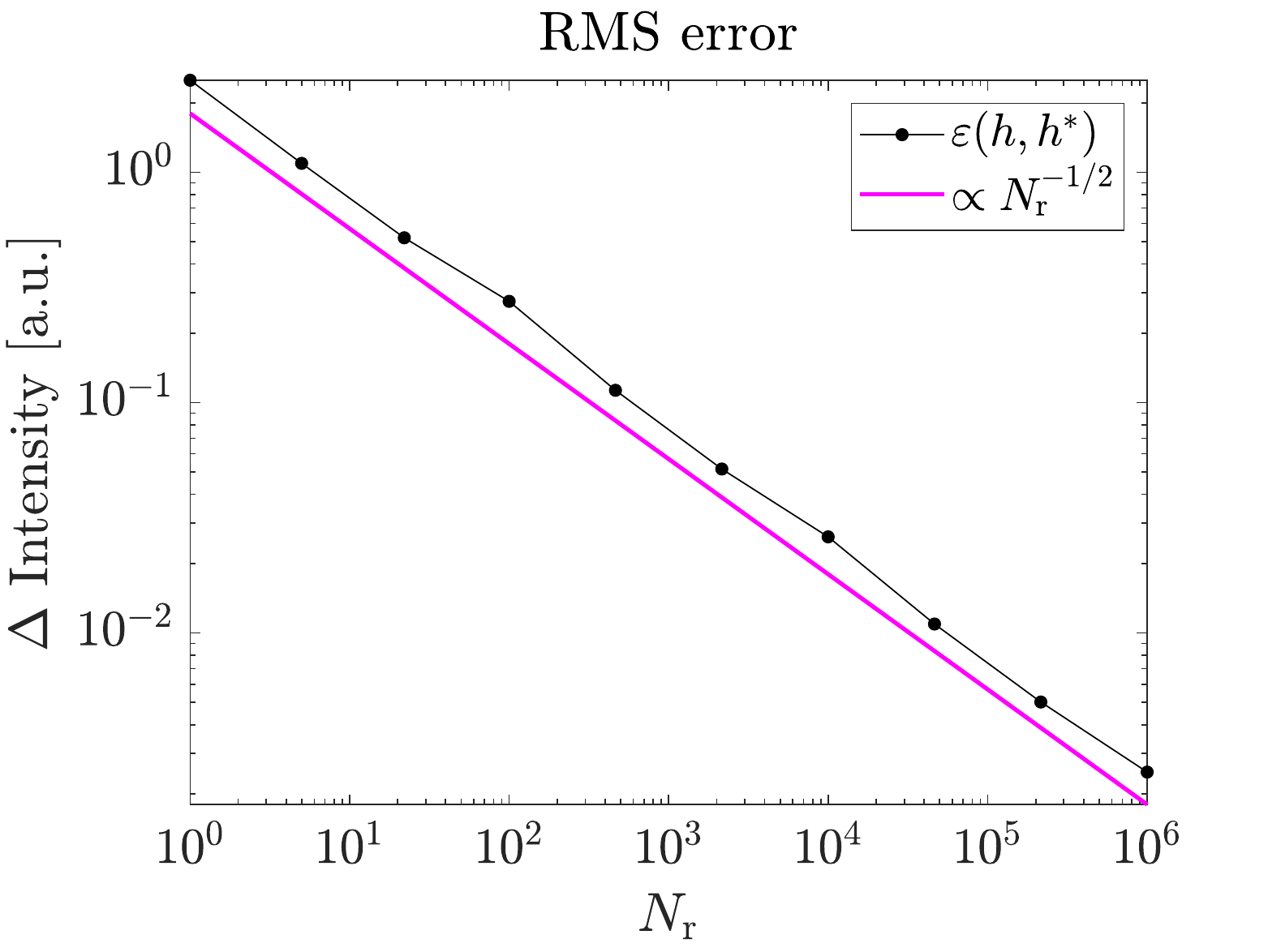}
	\captionsetup{width=\linewidth}
	\caption{Raytraced distributions in Example \#1; 256 sample points, $10^6$ rays traced.}
	\label{fig:example_1-3}
\end{figure}

We then raytraced these reflectors to validate our model of scattering.
This procedure was outlined in Sec.~\ref{sec:validation} and the result is shown in Fig.~\ref{fig:example_1-3} for the $\mathbf{r}_-$ reflector, computed using the monotonically decreasing optical map $m_-$.
Note the near-perfect $-1/2$ exponential convergence when the RMS error $\varepsilon(h,h^*)$ --- recall Eq.~\eqref{eq:h_RMS} --- is plotted against the number of rays traced, $N_\mathrm{r}$, as expected from Monte Carlo raytracing \cite[p.~9]{filosaPhaseSpaceRay2018}.
The only problematic areas in the raytraced distributions are those close to $\psi_1$ and $\psi_2$ for $g^*$.
This is smoothed out by the effect of scattering so that the scattered distribution is nearly ideal.
As this manuscript is focused on scattering in freeform reflector design, we are perfectly content with this minor discrepancy.

\subsection{Point Source}
Having verified our model, we shall now consider an example problem using a point source --- see the box below.
This differs from the previous example in two crucial ways, in addition to the apparent difference in symmetry going from a parallel source to a point source.
Namely, the exact $g$ is no longer known, as we only prescribe a desired scattered target distribution $h$.
In addition, the scattering function $p$ is now given by a Lorentzian (also known as a Cauchy distribution)
\begin{equation}
	p(\eta) = \mathcal{L}(\eta;\sigma) := \frac{1}{\pi\sigma}\Bigg(\frac{\sigma^2}{\eta^2+\sigma^2}\Bigg),
\end{equation}
with a full width at half maximum (FWHM) of $2\sigma$.
This change in $p$ is significant for two reasons: machined mirrors often exhibit this type of BRDF \cite[Ch.~4]{stoverOpticalScatteringMeasurement2012}, and the tails fall to zero at a significantly lower rate compared to the Gaussian we used in our previous example.

\begin{mdframed}
	\textbf{Example \#2: Unknown Specular Distribution}
	\begin{align*}
		&\text{$\th$-range:} 						&[\th_1, \th_2] 			&= [-0.47,1.41]\\
		&\text{$\gamma$-range:} 					&[\gamma_1,\gamma_2] 		&= [-1.41,1.02]\\
		&\text{$\eta$-range:}						&[\eta_1,\eta_2]			&= [-\pi/2,\pi/2]\\
		&\text{Source distribution:} 				&f(\th)						&= 1\\
		&\text{Scattered target distribution:} 		&h(\gamma) 					&= \cos^{16}(\gamma - 0.3) + 1.5\cos^8(\gamma + 0.4) - 0.01\\
		&\text{Surface scattering function:} 		&p(\eta) 					&= \mathcal{L}(\eta;5^{\circ})
	\end{align*}
\end{mdframed}

\begin{figure}[H]
	\centering
	\includegraphics[width=0.5\linewidth]{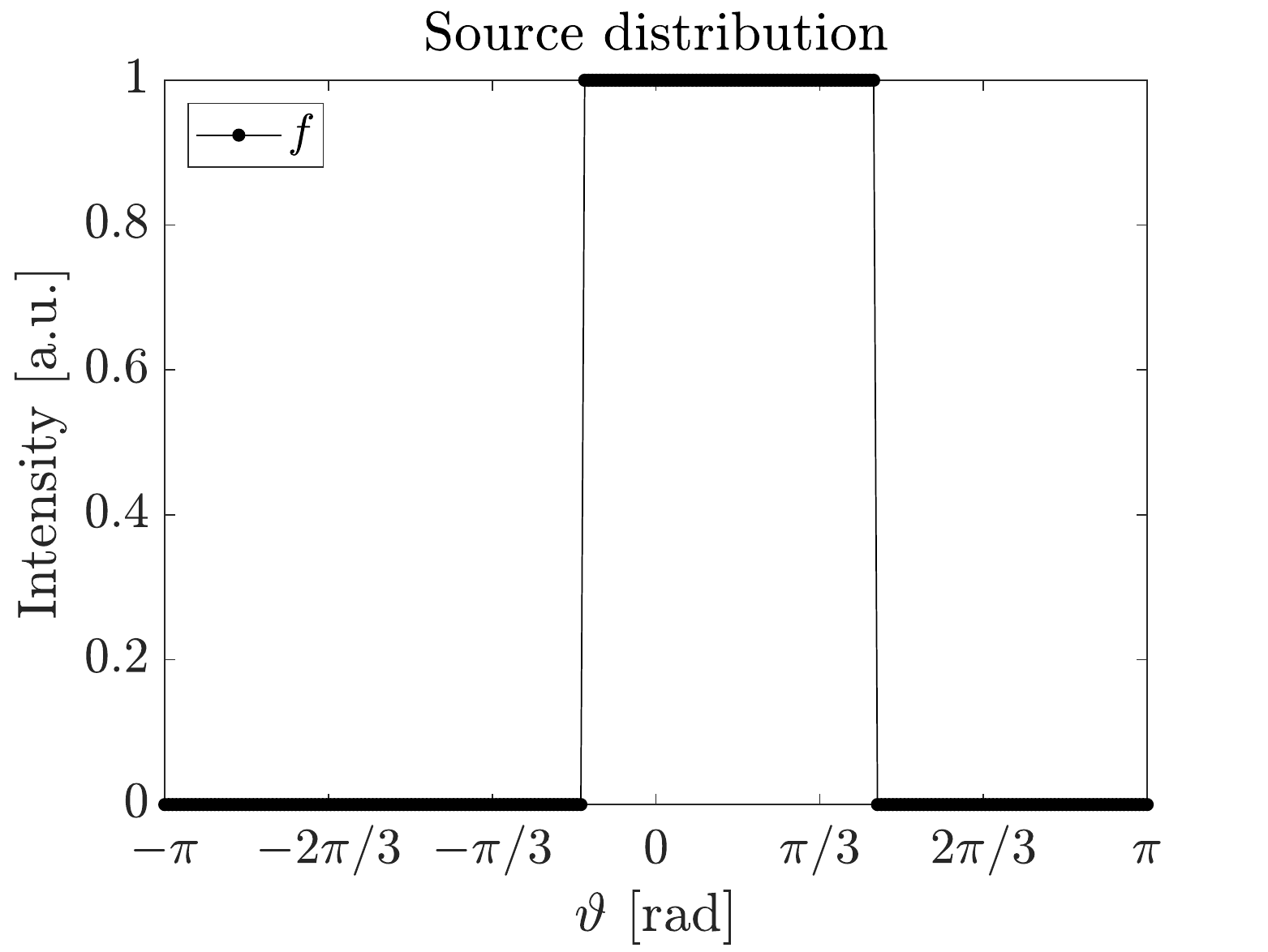}%
	\includegraphics[width=0.5\linewidth]{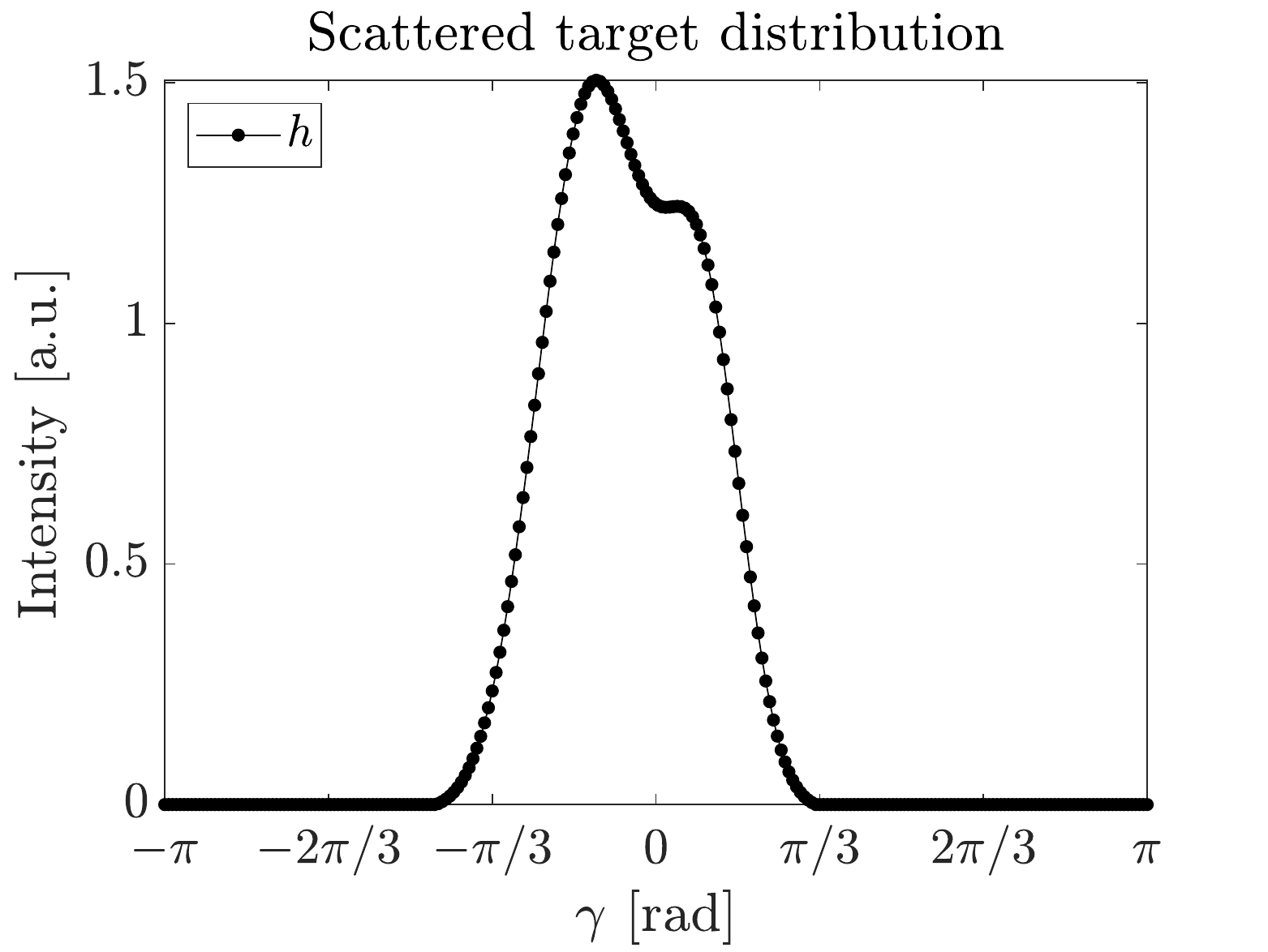}\\[5pt]
	\includegraphics[width=0.5\linewidth,page=2]{TikZ/spyFigs/spyFigs}
	\captionsetup{width=\linewidth}
	\caption{Distributions in Example \#2; $256$ sample points.}
	\label{fig:example_2-1}
\end{figure}

The distributions are shown in Fig.~\ref{fig:example_2-1}, were we note the sharp cutoff of the tails of $p$ at $\eta = -\pi/2$ and $\eta = \pi/2$.
The distributions underwent the same treatment as in the first example, i.e., clamping to finite domains and renormalisation as appropriate.

\begin{figure}[H]
	\centering
	\includegraphics[width=0.5\linewidth]{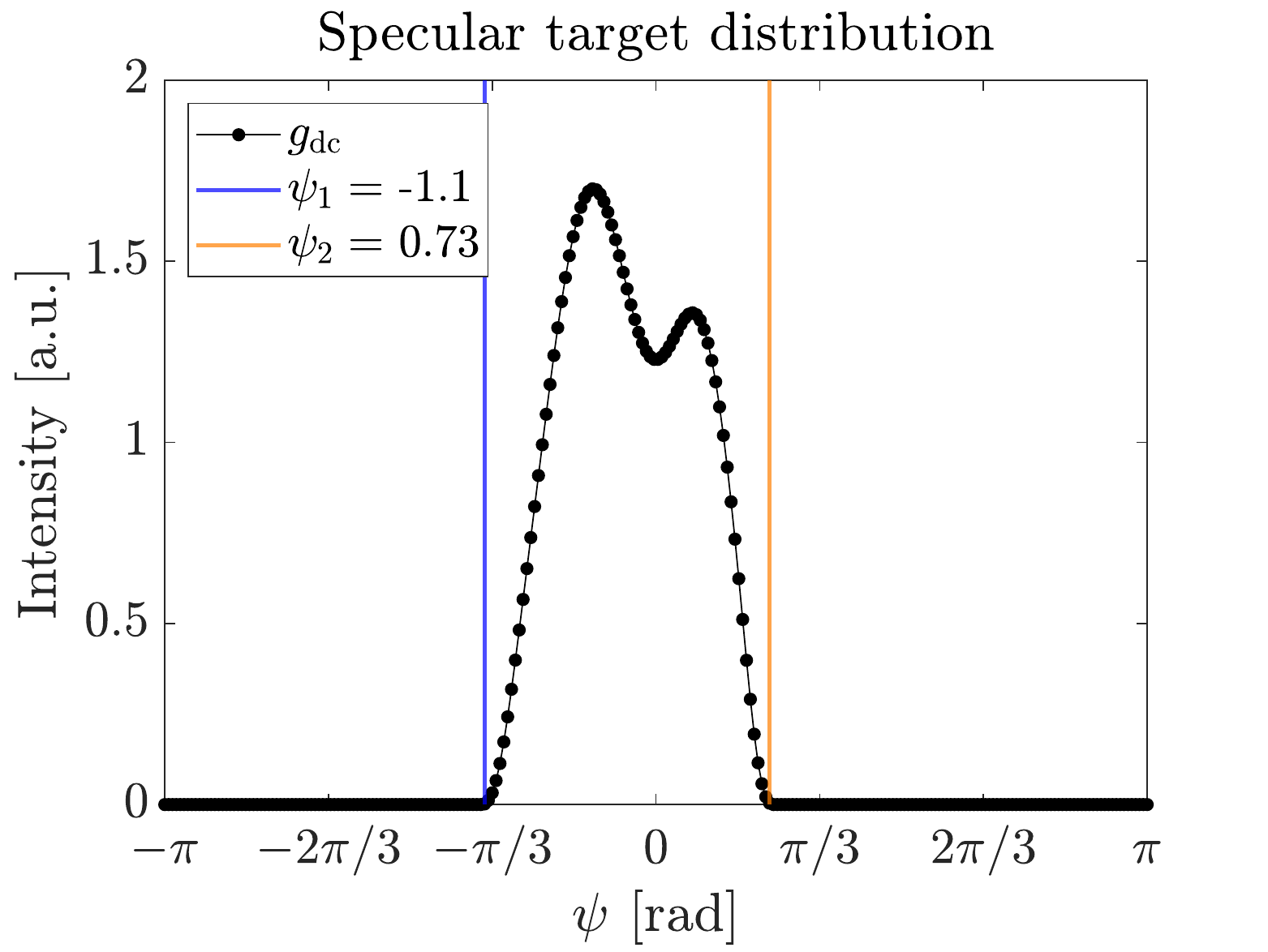}\\[5pt]
	\includegraphics[width=0.5\linewidth]{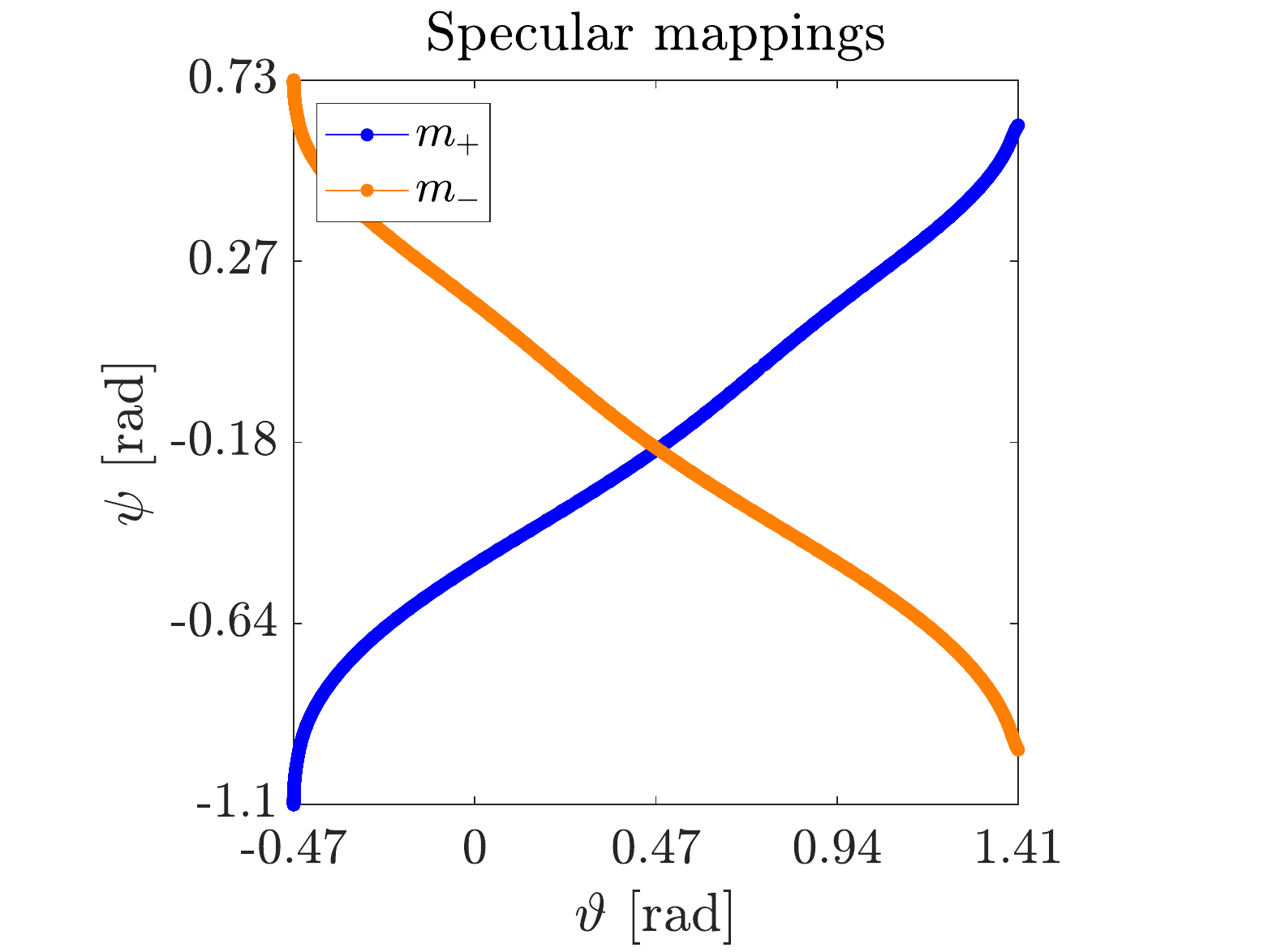}%
	\includegraphics[width=0.5\linewidth]{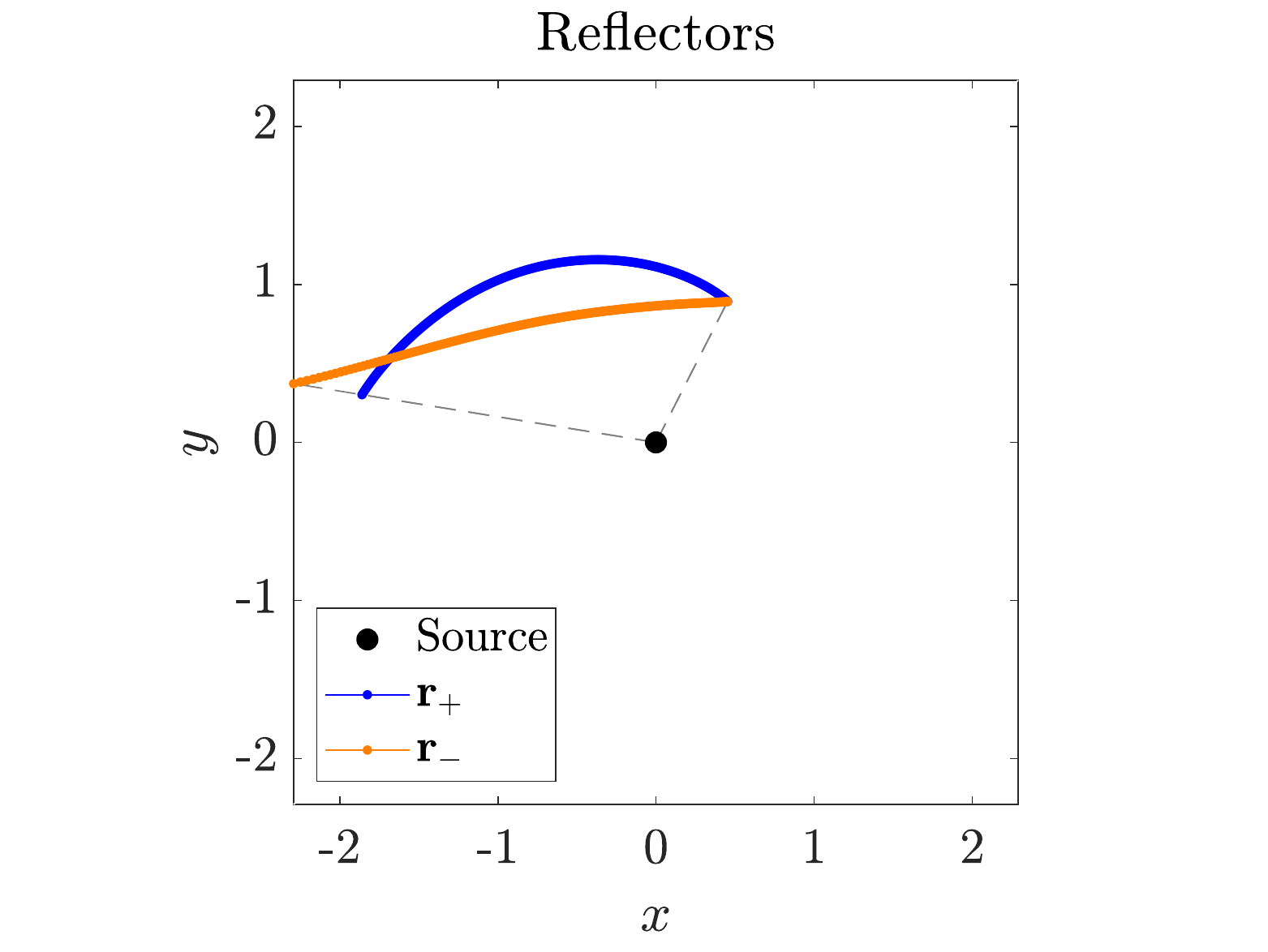}
	\captionsetup{width=\linewidth}
	\caption{The finite domain of $g_\mathrm{dc}$, specular mappings $m_\pm$ and associated reflectors $\mathbf{r}_\pm$ in Example \#2; 256 sample points of $g_\mathrm{dc}$ and $\mathbf{r}_\pm$.}
	\label{fig:example_2-2}
\end{figure}

To compute the reflectors, we first determined a deconvolved specular target distribution $g_\mathrm{dc}$ using ten iterations of Matlab's \texttt{dconvlucy} deconvolver, implementing the Richardson-Lucy deconvolution scheme.
We then solved the IVPs for the optical mappings $m_\pm$ as well as those for the reflector radius functions $u_\pm$.
This is shown in Fig.~\ref{fig:example_2-2}, together with the limits $\psi_1$ and $\psi_2$ (found via $g(\psi) = \epsilon$) used to solve the IVPs for $m_\pm$ --- recall Eqs.~\eqref{eq:opticalMap_pointPlus} and \eqref{eq:opticalMap_pointMinus}.
The IVPs for $m_\pm$ and $u_\pm$ were solved in the same way as in the previous example, i.e., using the ODE solver \texttt{ode15s} with improved tolerances and by solving the equivalent integral equation, Eq.~\eqref{eq:IVPSolPoint}, using Matlab's \texttt{integral} command, respectively.

\begin{figure}[H]
	\centering
	\includegraphics[width=0.5\linewidth]{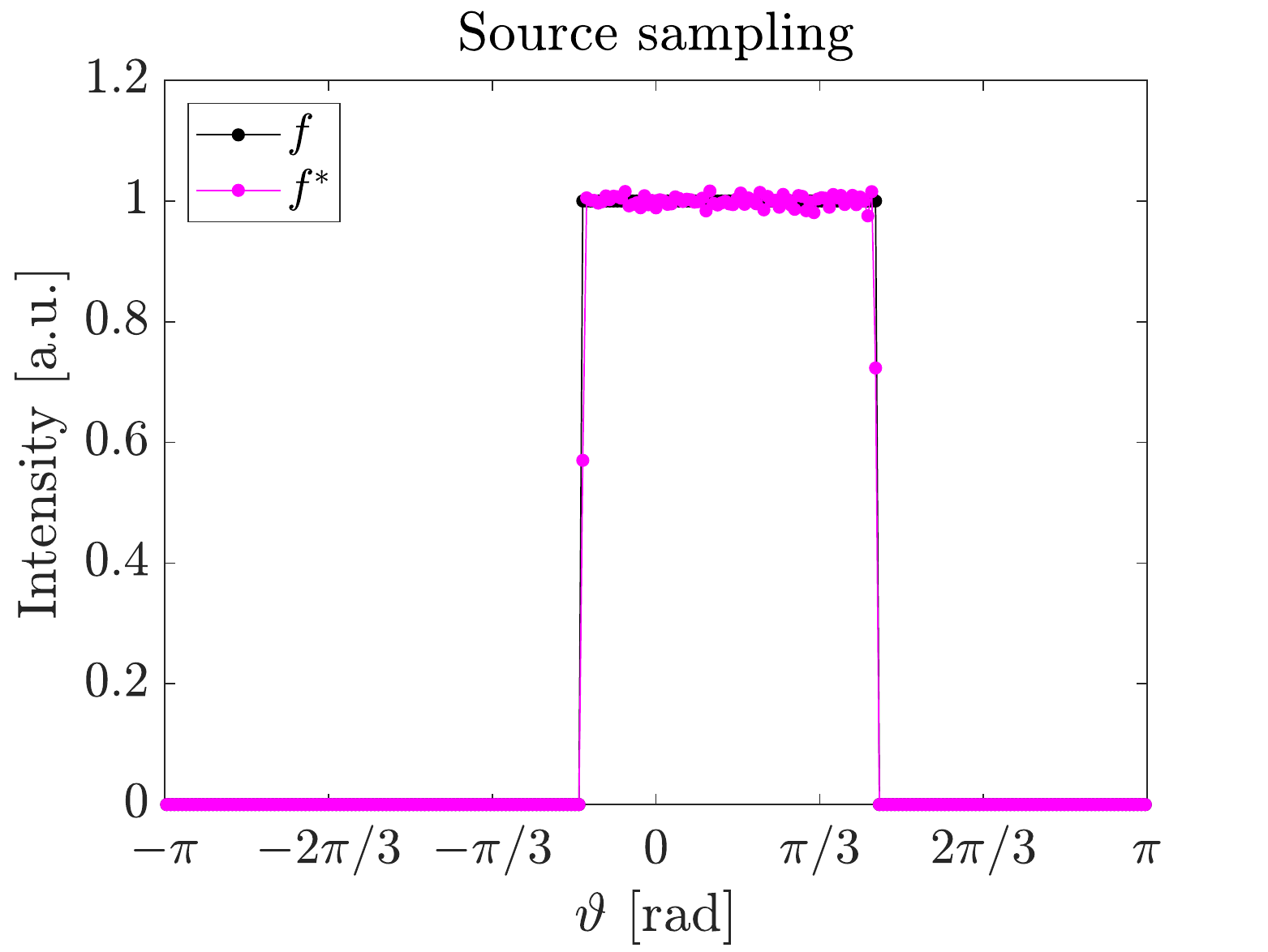}%
	\includegraphics[width=0.5\linewidth,page=3]{TikZ/spyFigs/spyFigs}\\[5pt]
	\includegraphics[width=0.5\linewidth]{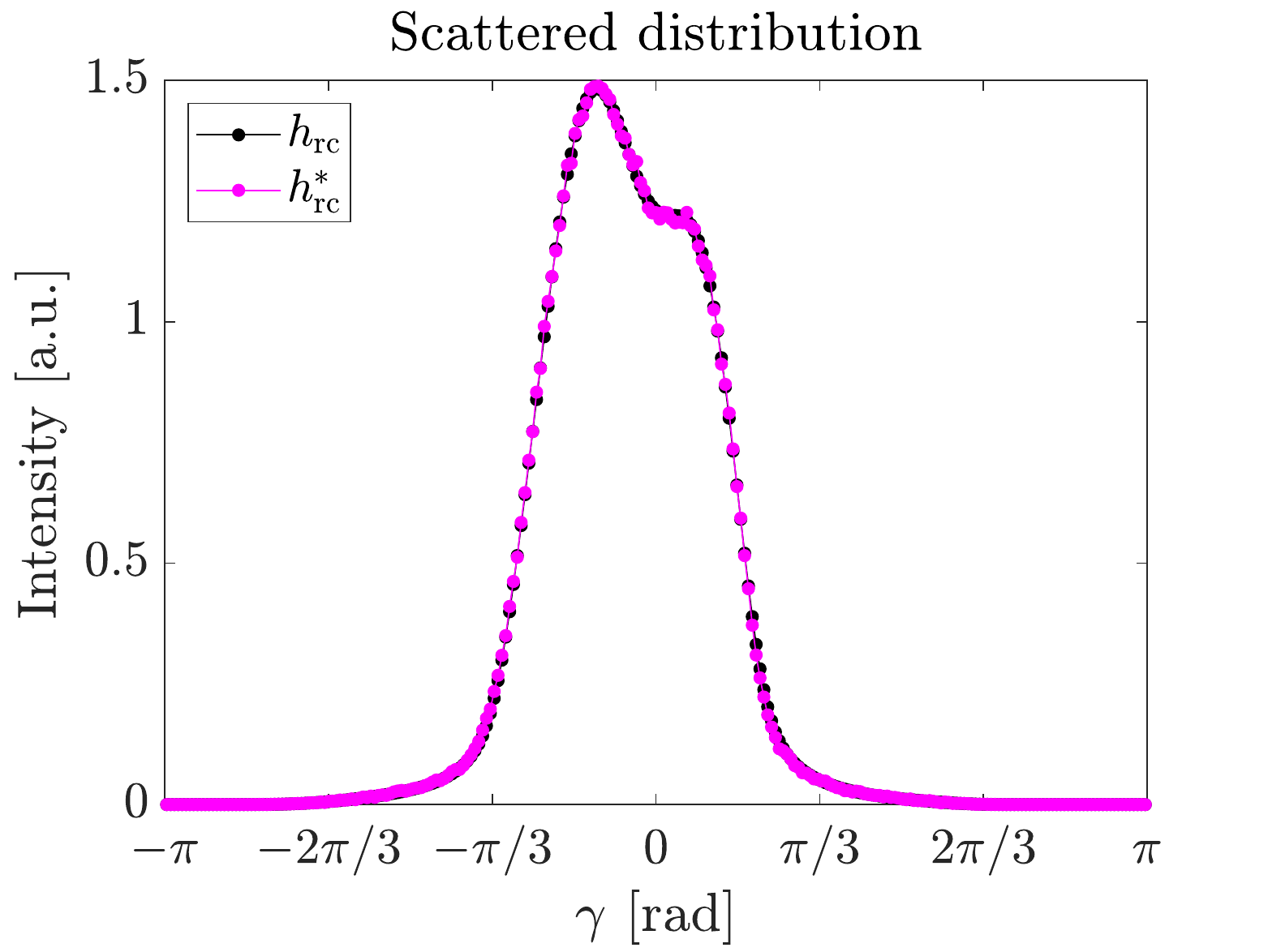}%
	\includegraphics[width=0.5\linewidth]{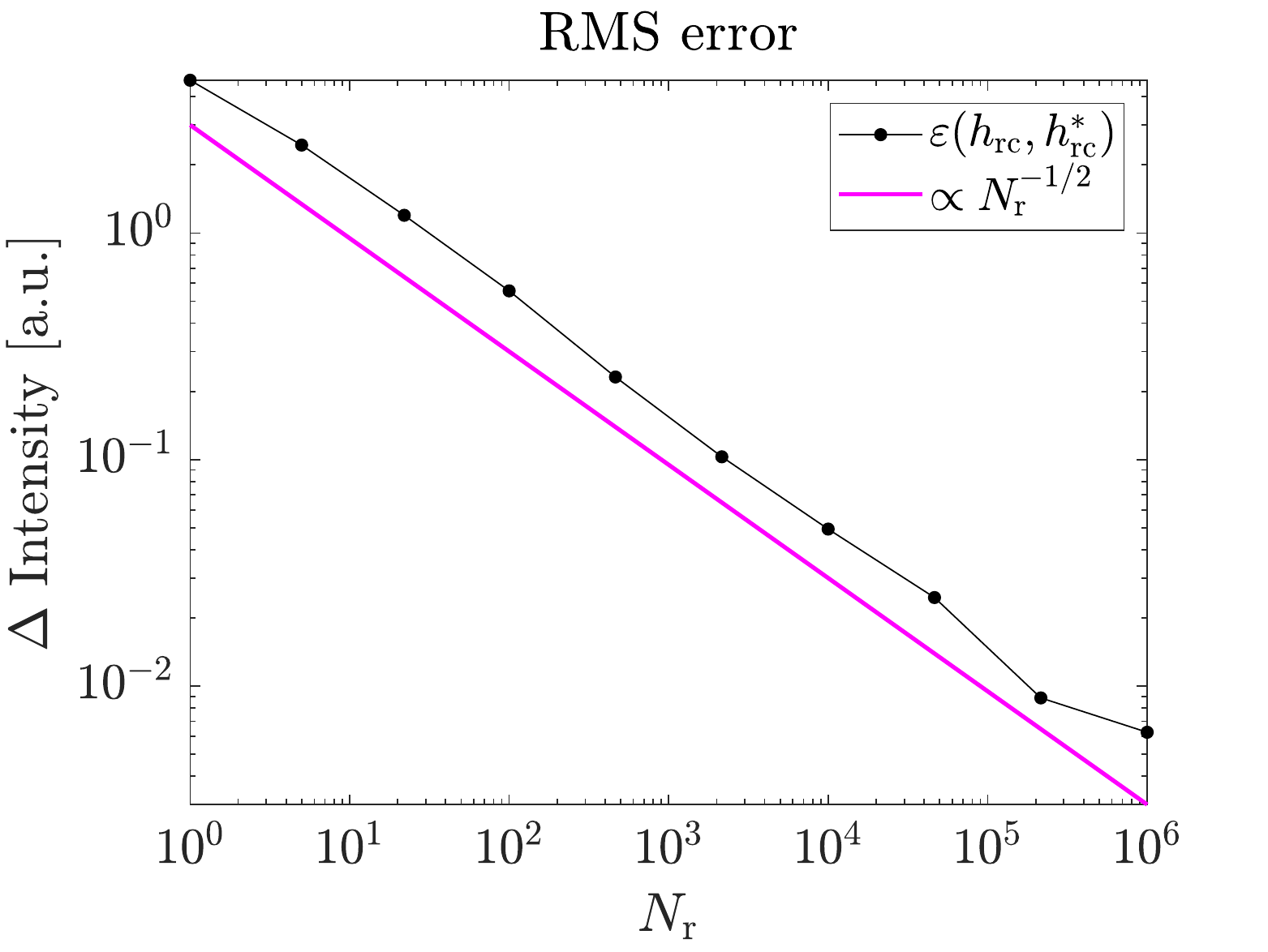}\\[5pt]
	\includegraphics[width=0.5\linewidth]{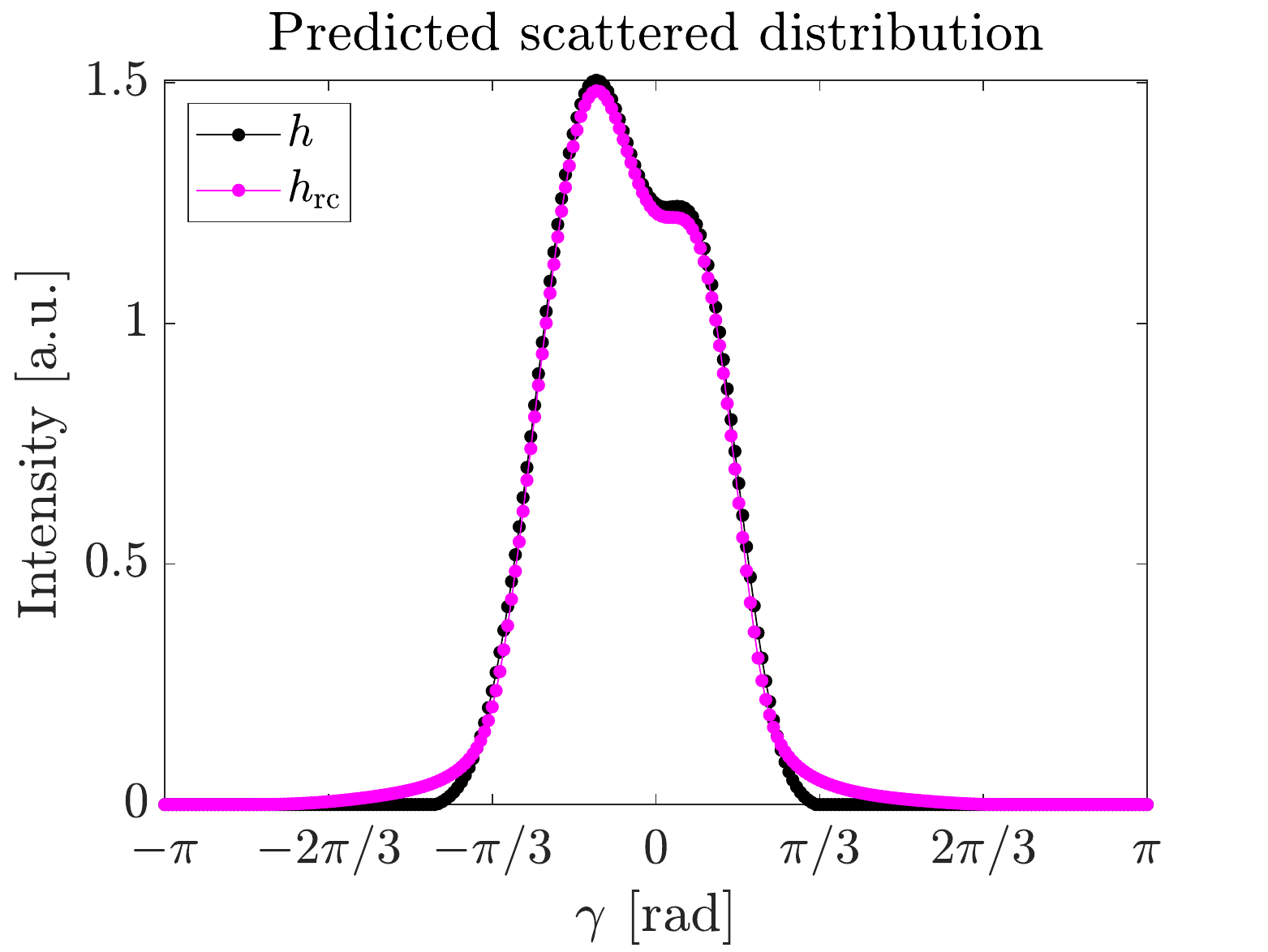}
	\captionsetup{width=\linewidth}
	\caption{Raytraced distributions, $h$ and $h_\mathrm{rc}$ in Example \#2; 256 sample points, $10^6$ rays traced.}
	\label{fig:example_2-3}
\end{figure}

We then raytraced the $\mathbf{r}_+$ reflector to obtain the data in Fig.~\ref{fig:example_2-3}, where we again observe the expected $-1/2$ trend with the number of rays traced.
Looking at the specular distribution, we notice that the behaviour close to the boundaries $\psi_1$ and $\psi_2$ is slightly worse than in the previous example.
Given the fact that deconvolution is an ill-posed problem \cite[Ch.~1, Sec.~V, p.~32]{janssonDeconvolutionImagesSpectra2012}, it is not obvious that the model would work well for this example.
Indeed, by prescribing only the scattered distribution $h$, rather than the specular $g$, we are not guaranteed existence or uniqueness of $g_\mathrm{dc}$.
Numerically, we can still apply Matlab's \texttt{deconvolucy} iterative deconvolver to find \textit{a} deconvolved distribution $g_\mathrm{dc}$, of course, but the existence of a \textit{true} $g$ is not a given.
This is also why we compare the raytraced scattered distribution to $h_\mathrm{rc} := p*g_\mathrm{dc}$, i.e., the `reconvolved' distribution, rather than the prescribed $h$.
The difference between the two is shown at the bottom of Fig.~\ref{fig:example_2-3}.

\clearpage
\section{Conclusions}\label{sec:conclusion}

\noindent We have presented a general method of computing two-dimensional (and by extension rotationally and cylindrically symmetric) scattering reflector surfaces.
Specifically, by considering all the scattered light from a curve endowed with microfacets, a convolution integral between a probability density function (PDF) related to the stochastic orientation of the microfacets --- and thus to surface roughness --- and a virtual specular target distribution representing the reflected light from a smooth surface, was reached.
Deconvolution thus yields an approximation of the virtual specular target distribution after prescribing a desired scattered target light distribution and the PDF.
In this way, the problem is reduced to a well-understood specular inverse problem.
In its current form, the main strength of our approach is thus that it can be performed as a pre-processing step before using already mature specular design workflows.

The model was verified by comparing the predicted light distribution from the convolution integral with one obtained using a custom raytracer, which directly implemented the microfacets used as a basis for our model, with reflectors computed using the deconvolved virtual specular target distribution.
This verification step showed that our model was consistent for parallel and point sources.
Due to the ill-posedness of deconvolution, the proposed algorithm is most effective for smooth scattered target distributions, and discontinuous scattered target distributions cannot be deconvolved since no sharper feature can be created.

This work makes it easy to think of many exciting extensions.
Some examples include treating refractive media (i.e., lenses), three-dimensional freeform reflectors, and allowing the PDF related to the roughness of the surface to change as a function of position or to take into account effects such as shadowing and masking, i.e., the idea that adjacent microfacet may obstruct the flux incident or scattered from a given microfacet.
In the case of refractive media, one could, for instance, use our model to correct for surface scattering due to machining imperfections.
Allowing the PDF to change along the two-dimensional reflector results in a type of Fredholm integral equation.
We are actively investigating every aforementioned extension.\\

\hrule
\begin{changemargin}{1cm}{1cm}
	\noindent\textsc{\textbf{Funding:}} This work was partially supported by the Dutch Research Council (\textit{Dutch:} Nederlandse Organisatie voor Wetenschappelijk Onderzoek (NWO)) through grant P15-36.\\
	\textsc{\textbf{Disclosures:}} The authors declare no conflicts of interest.\\
	\textsc{\textbf{Data availability:}} Data underlying the results presented in this paper are not publicly available at this time but may be obtained from the authors upon request.
\end{changemargin}
\hrule

\pagestyle{ref}
\addcontentsline{toc}{section}{References}
\printbibliography

\end{document}